\documentclass[twocolumn,showkeys,showpacs,preprintnumbers,amsmath,amssymb,aps,prb,floatfix,longbibliography,superscriptaddress,floatfix,10pt]{revtex4-2}
\usepackage[skip=2pt]{caption}

\usepackage{float} %Gráficas y tablas
\usepackage{verbatim} %Multiples comentarios
\usepackage[skip=2pt]{caption}

\usepackage{lipsum,capt-of,graphicx} %Gráficas
\usepackage{subfigure} %Gráficas
\usepackage{float} %Gráficas y tablas

\usepackage{graphicx}% Include figure files
\usepackage{dcolumn}% Align table columns on decimal point
\usepackage{bm}% bold math
\usepackage{dcolumn}% Align table columns on decimal point
\usepackage{bm}% bold math
\usepackage[usenames]{color}
\usepackage[%
  colorlinks=true,
  urlcolor=blue,
  linkcolor=blue,
  citecolor=blue
]{hyperref}
\usepackage{pifont}
\usepackage{xcolor} 
\usepackage[utf8]{inputenc}
 \usepackage{ulem}

\begin{document}

\title{Tax evasion study in a society realized as a diluted Ising model with competing interactions}
\
\author{Julian Giraldo-Barreto}

\email{julian.giraldob@udea.edu.co}

\affiliation{Magnetism and Simulation Group G+, Institute of Physics, University of Antioquia. A.A. 1226, Medell\'in, Colombia}

\affiliation{Biophysics of Tropical Diseases Max Planck Tandem Group, University of Antioquia. A.A. 1226, Medell\'in, Colombia}

\author{J. Restrepo}

\email{johans.restrepo$@$udea.edu.co}

\affiliation{Magnetism and Simulation Group G+, Institute of Physics, University of Antioquia. A.A. 1226, Medell\'in, Colombia}

\thispagestyle{empty}

%\date{\today}

\begin{abstract}
In this research, the tax evasion percentage, as order parameter, of a system of individuals or agents inscribed in a $N = L \times L$ 2D square grid is computed. The influence of local environment over each agent is quantified both through competitive exchange integrals (ferromagnetic and antiferromagnetic bonds) and dangling bonds randomly distributed, which allows to identify the system with disordered ternary alloys of the type $\mathrm{A_\textit{p}B_\textit{x}C_\textit{q}}$ with a certain stoichiometry $(p,x,q)$ particular of each society. Our proposal is based on the so-called spin glass phase present in magnetic systems characterized by disorder, dilution and competitive interactions where magnetic frustration can take place, resembling the way as an individual or agent in a society is able to face a decision. In this sense, agents are identified as Ising spins, which can take two possible values ($\sigma = \pm 1$), in correspondence with a two-state system where agents can be tax compliant or not. Such an identification between social and physical variables, as well as parameters like an external applied magnetic field or temperature, are topic of discussion in this investigation. Thermalization of the observables is carried out by means of the heat bath algorithm. Other social variables, such as the audit period, and its effects over the percentage of evasion, are used to analyze the behavior of tax evasion in Colombia, however the model can be applied to any country.
\end{abstract}


\maketitle

\section{Introduction}

Tax evasion is an issue that concerns the global economy, since tax collection is a first step in the economic growth of a country. However, the great benefits that an efficient tax system can bring to a nation can be overshadowed by the systematic abuse of tax collection or corruption\cite{Teymur2012}, the feeling of getting rich, or the simple fact of not understanding the law, which inevitably leads to a sector of the population to become evader of its tax obligation, thus contributing to weaken the economy of a country\cite{Lunina2020}. Thus, tax evasion arises naturally as a field of study that must be tackled from a wide range of disciplines that allow a clearer knowledge of its most influential variables, so that in conjunction with the dynamics of government and society in general, this misconduct is minimized.

The first proposal can be traced to the work of Allingham \& Sandmo \cite{Allingham1972,Yitzhaki1974}. Such study dealt with the expected value of the profit obtained by an agent depending on the income declared, and the fraction of taxes to be paid according to what has been declared. This work also investigates the conditions dealing with the maximization of the profit, finding that, declaring less income than actually earned (which is nothing more than a form of evasion), is a sufficient condition to maximize such utility.

From the experimental point of view, Bosco \& Mittone~\cite{Bosco1997}, and more recently Mittone~\cite{Mittone2006} focused on how psychological factors influence tax evasion. In essence, it follows that a redistribution of the tax yield reduces tax evasion, and the way the agent perceives the detection and punishment risk becomes also determinant~\cite{Mittone2006}. Such psychological factors tend to be correlated with the way the audit is done over the tax compliance population.

With the improve of computing features and algorithms, agent-based models\cite{Hokamp2010a,Garrido2013,Hokamp2014,Koehler2018,GarciaAlvarado2019} become an effective tool for describing the tax evasion of a society, including social parameters that take into account the possible scenarios in which a group of agents susceptible to declare taxes are involved.

Zaklan and Zaklan \& Lima~\cite{Zaklan2008,Lima2008,Zaklan2009} used an Ising model~\cite{Ising1925} to describe a society composed of agents who can be compliant or non-compliant tax payers, applying social variables such as penalty periods and an audit probability within their study. In the context of an Ising model, it is assumed that agents that do not evade (the compliant ones) are identified with particles whose spin values are $\sigma_i = +1$, while the evaders (the non-compliant ones) are identified with spin values equal to $\sigma_i = -1$. Lima \cite{Lima2012a,Lima2012,Lima2015,Lima2015a} extend Zakland's work \cite{Zaklan2009} by investigating characteristics like phase transitions, and different networks in which agents lie in the framework of a pure Ising model.

The above opened a new range of possibilities: in the works of Seibold \& Pickhardt \cite{Seibold2013b,Pickhardt2014a} they used physical variables in the framework of an Ising model to study the percentage of evasion of a system of agents. Such system was divided into 4 groups of agents according to their social conduct, namely: (i) totally honest or ethical agents, (ii) totally evading or selfish agents, (iii) agents influenced by their local environment, the so-called “copying" agents, and (iv) random agents who act by chance. Such classification was parameterized in terms of the interplay between temperature, magnetic field and exchange integral. In particular, both the magnetic field and temperature were considered as local variables whose values were taken different from site to site in the grid, dealing, in the case of temperature, with the personality of the agents and therefore with the degree of autonomy to make a decision. Their studies reproduced results of agent-based models that incorporate characteristics of the study made by Allingham \& Sandmo \cite{Allingham1972}. This makes clear that, the models that are identified in the current literature as econophysical models, manage to extract the essentials of purely economic models, obtaining comparable results and making them more robust and accessible to a diverse scientific community\cite{Bazart2016}. For example, one of the latest works of Berger and collaborators~\cite{Berger2020}, studied the ``bomb crater effect" by means of a heterogeneous Ising model, following the same conditions as~\cite{Seibold2013b} or~\cite{Pickhardt2014a}, about particularizing the magnetic field and the temperature as different for each agent.

As concerns to a case study, concretely for the Republic of Colombia, characterized historically by a high degree of tax evasion, a review of the state of the art has only been addressed by administrators or economists in recent years. The works of Parra Jimenez \& Patiño Jacinto \cite{Jimenez2010} or Rodríguez Rodriguez-Cuervo \cite{Cuervo2018} seek to analyze the tax evasion rates for the periods 2$001-2009$ and $1997-2017$, respectively. They studied the possible causes of tax evasion and some alternatives to reduce it, besides estimates of the percentage of evasion were also given. This has only been seen from a purely economic framework, wherewith the application of a general model to this particular case could give future perspectives to reduce, or at least control, tax evasion in Colombia. In such works, evasion rates based on official data obtained from the National Tax and Customs Directorate (DIAN), are presented, and therefore they are elements to be considered in this investigation.

The manuscript is organized as follows: In Sec. \ref{Model_Sec}, we present the model, the physical insights involved in the simulations and the equivalence between physical and social variables. In particular, we show how the use of competitive exchange integrals and diluted bonds randomly distributed in the framework of a spin glass-type Ising model, allows to identify our system of agents like a disordered ternary alloy where stoichiometry plays a key role. In Sec. \ref{Method_Sec} methodology and computational details including the heat bath algorithm are presented. In Sec. \ref{Results_Sec} we analyze the effect of auditing and government policies for tax collection, the role that temperature plays in our approach as well as the exchange integrals, and a final analysis of application of our model for the Republic of Colombia. We finish with a discussion about the relevance of these results and we present the main conclusions. 

\section{Model}
\label{Model_Sec}

The two-state character of compliance of the agents $\sigma_i$ is modeled by considering two possible values, namely: (i) a positive one corresponding to a compliant tax payer with spin $\sigma_i=+1$ and (ii) a negative one for a non-compliant tax payer with spin $\sigma_i=-1$. On the other hand, we propose that instead of a predetermined  agent classification in terms of personality, autonomy or even degree of selfishness, a range of subjective scenarios can arise in a natural way if the lattice of agents is realized as a diluted Ising system with competing interactions, which are features of a spin glass in solid state physics. Such a system is characterized by a random distribution of nearest neighbors competitive ferromagnetic ($J_{ij}>0$) and antiferromagnetic ($J_{ij}<0$) bonds and also by a random distribution of dangling or diluted bonds ($J_{ij}=0$). In this way, a distribution of magnetic coordination numbers, in conjunction with the present interactions, can lead spontaneously to an agent frustration at a given lattice site, similar to magnetic frustration in spin glass systems. Thus, frustration, related to human behavior when faced with the need to make a decision, can arise naturally depending on the local environment surrounding a given individual. Based on this hypothesis, competitive bonds are responsible for the contradictory or confusing information to which an individual may be exposed whereas different local concentration of diluted bonds lead to a diverse degree of autonomy of loneliness when facing a decision about compliance. Consistently with the arguments exposed above, our diluted Ising Hamiltonian reads as follows:
\begin{equation}
\begin{split}
 \cal{H}=&-\sum_{\langle i,j \rangle } J_{ij}\sigma_i\sigma_j -\sum_{i} \sigma_i H+\\
 & \sum_{i\in A|\sigma_i=-1}\sigma_ig_i(t-h),
 \end{split}
\label{eqn:eqn1}
\end{equation}
where the first sum runs over nearest neighbors $\langle i,j \rangle$, and where exchange integrals $J_{ij}$ follow a probability distribution function (p.d.f) including competitive and diluted bonds. A way to accomplish this is by considering our system of agents like a ternary disordered alloy of the form $\mathrm{A_\textit{p}B_\textit{x}C_\textit{q}}$ with a certain stoichiometry $(p,x,q)$, particular of each society, with $p+x+q=1$. In this context, the p.d.f can be written as:
\begin{equation}
\begin{split}
P(J_{ij})=& p^2\delta(J_{ij}-J) + x^2\delta(J_{ij}+J) + 2px\delta(J_{ij}+J) +\\
& (q^2+2pq+2qx)\delta(J_{ij})
\end{split}
\label{eqn:eqn2}
\end{equation}
In this expression $p^2$ is the probability of ferromagnetic $\mathrm{AA}$ bonds with $J_{ij}=J>0$, $x^2$ and $2px$ are the probabilities of antiferromagnetic $\mathrm{BB}$ and $\mathrm{AB}$ bonds respectively with $J_{ij}=-J<0$, and $q^2$, $2pq$, and $2qx$ are the respective probabilities of diluted bonds involving a dilutor element $\mathrm{C}$, namely: $\mathrm{CC}$, $\mathrm{AC}$ and $\mathrm{BC}$ with $J_{ij}=0$. In this work, such an implementation has been carried out by introducing a homogeneous Python3.x \cite{VanRossum2009} uniform function \texttt{random.choice(-$J$,0,$J$)}. This election implies an \textit{a priori} equiprobability assumption for which $P(-J)=P(0)=P(J)=1/3$ or, in other words:
\begin{equation}
p^2 = x^2 + 2px = q^2+2pq+2qx = 1/3,
\label{eqn:eqn3}
\end{equation}
which allows to define the stoichiometry for a homogeneous case resulting in $p=1/\sqrt{3}\approx 0.577$, $x=(\sqrt{2}-1)/\sqrt{3}\approx 0.239$ and $q=(\sqrt{3}-\sqrt{2})/\sqrt{3}\approx 0.184$. In such a homogeneous case, which is our concern in this work as a first approximation, the dilutor element does not constitute a real agent in the society, but the absence of a real one, which allows to span different degrees of autonomy for the actual agents in the system. On the other hand, the other integrals are assumed to be $J_{AA}=J$ and $J_{AB}=J_{BB}=-J$. From here, it is clear that different values of $J$ can in principle be considered in order to account for different degrees of strength in the coupling or influence of an agent with its local environment. Additionally, other different stoichiometries $(p,x,q)$, corresponding to non homogeneous cases, can in principle give account of different societies.

Both, the second and third terms in equation \ref{eqn:eqn1} stand for government interventions and they are Zeeman-type interactions of the agents (or spins). Concretely, the second term gives the interactions of all the agents with a global external field $\vec{H}$ dealing with the general government policies, which are the same for everyone, although it could also be related with the image of favorability of a government. The third term is stochastic, and it gives the interaction of a subset $A$ of non-compliant tax payers ($\sigma_i=-1$) with a local audit field $g(h)$ relative to a random intervention of the state, which in turn depends on a variable $h$ accounting for the number of periods of penalization to be applied \cite{Zaklan2008,Zaklan2009, Seibold2013b,Pickhardt2014a}. The random character of this last interaction implies also a probability of the form:
\begin{equation}
P(g_i(t-h)) = \alpha\delta(\sigma_i +1),
\label{eqn:eqn4}
\end{equation}
where $\alpha$ is the probability of audit intervention. Finally, $g_i(t-h)$ is a Heaviside function that forces the non-compliant tax payer to be honest, i.e. $\sigma_i=-1 \to +1$, within a penalization period $h$. Hence,
\begin{equation}
g_i(t-h)= \left\{ \begin{array}{lcc}
             -\sigma_i &   if  & 0 \leq t \leq h \\
             \\ 0 &  if  & t > h
             \end{array}
   \right.
\label{eqn:eqn5}
\end{equation}

\section{Methodology and Computational details}
\label{Method_Sec}

Thermalization has been conducted by using the heat-bath algorithm~\cite{Krauth2006,Jedrzejewski2017} where the probability $\pi(\sigma_i)$ of an agent $\sigma_i$ to take on the values $\pm 1$ in a thermal reservoir at an absolute temperature, $T$ (in units of $J/k_{B}$), is given by:
\begin{equation}
\pi(\sigma_i)=\frac{1}{1+\exp{-[E(-\sigma_i)-E(\sigma_i)]/k_{B}T}},
\label{eqn:eqn6}
\end{equation}
where $E(-\sigma_i)-E(\sigma_i)$ is the energy change associated to a spin-flip at site $i$ and one time step per spin or per agent. A full sweep implies trying an inversion in every single agent of the system with the probability $\pi(\sigma_i)$. In this work, up to $10^5$ sweeps were employed and up to $8\times 10^4$ sweeps, in regions with critical slowing down, were discarded for thermalization purposes. In general, the results in ~\cite{Jedrzejewski2017} show that using the heath bath algorithm works as good as the Metropolis one in the case the network is considered as a fully connected graph in the energy landscape. On the other hand, previous works ~\cite{Berger2020} ~\cite{Pickhardt2014a} ~\cite{Seibold2013b} show some results that are in agreement with studies done before by Zackland~\cite{Zaklan2008,Zaklan2009}, and others, by using the heath bath algorithm. It is important to stress at this point that the lattice is considered here in the framework of a canonical ensemble where the system is in thermal contact with a heat reservoir at an absolute temperature $T$, so temperature can not be regarded as a local parameter. This perspective differs from what has been done in other works where temperature has been considered as a local variable $T_i$ depending on the lattice site and dealing with the different types of agents conduct or personality \cite{Seibold2013b}. Moreover, in our case, temperature in addition to being considered as a global parameter, the one characterizing the bath, it must be related to the elapsed time required for a given policy to be adhered to and successfully completed. In this sense, the presence of any transition temperature $T_c$, can be ascribed to a deadline for paying taxes, which is reasonable to exhibit critical or pseudo-critical temperature characteristics. Thus, if we define the amount:
\begin{equation}
T_c-T=\Delta\tau,
\label{eqn:eqn7}
\end{equation}
we can warn that if the critical temperature is exceeded, there is a “negative time”, interpreted as an overdue time from a social perspective, and it will be in that time where evasion will play a crucial role.

Simulations were performed on a square lattice with periodic boundary conditions and a system size $N = L \times L$ agents with $L=100$ where finite size effects are negligible. Reduced units were employed. Energy values are the eigenvalues of the Hamiltonian shown in equation \ref{eqn:eqn1} whereas the order parameter may be related to either global magnetization or the fractional evasion. The former can be expressed, as is usual for magnetic systems, as:
\begin{equation}
m_k = \frac{1}{N}\sum_{i}\sigma_i,
\label{eqn:eqn8}
\end{equation}
and its corresponding thermal average is given by:
\begin{equation}
\langle |m|\rangle=\frac{1}{M_{max}-M_0}\sum_{k=M_0 +1}^{M_{max}}\ |m_k|,
\label{eqn:eqn9}
\end{equation}
where $M_{max}$ the maximum number of lattice sweeps and $M_0$ the cutoff value for thermal equilibration. On the other hand, fractional evasion (\textit{f.ev}) per sweep can be interpreted as a sublattice magnetization determined only by those non-compliant agents (this definition is similar to the used in \cite{Seibold2013b,Pickhardt2014a} ) ($\sigma_i=-1$), i.e.:
\begin{equation}
\textit{f.ev} = \frac{1}{N}\sum_{i|\sigma_i=-1}|\sigma_i|,
\label{eqn:eqn10}
\end{equation}
whence the respective average can be calculated similarly to equation \ref{eqn:eqn9}.

\section{Results}
\label{Results_Sec}

\subsection{Auditing and government policies for tax collection}

\begin{figure}
    %\justifying
    \includegraphics[width=8cm,
    height=7cm]{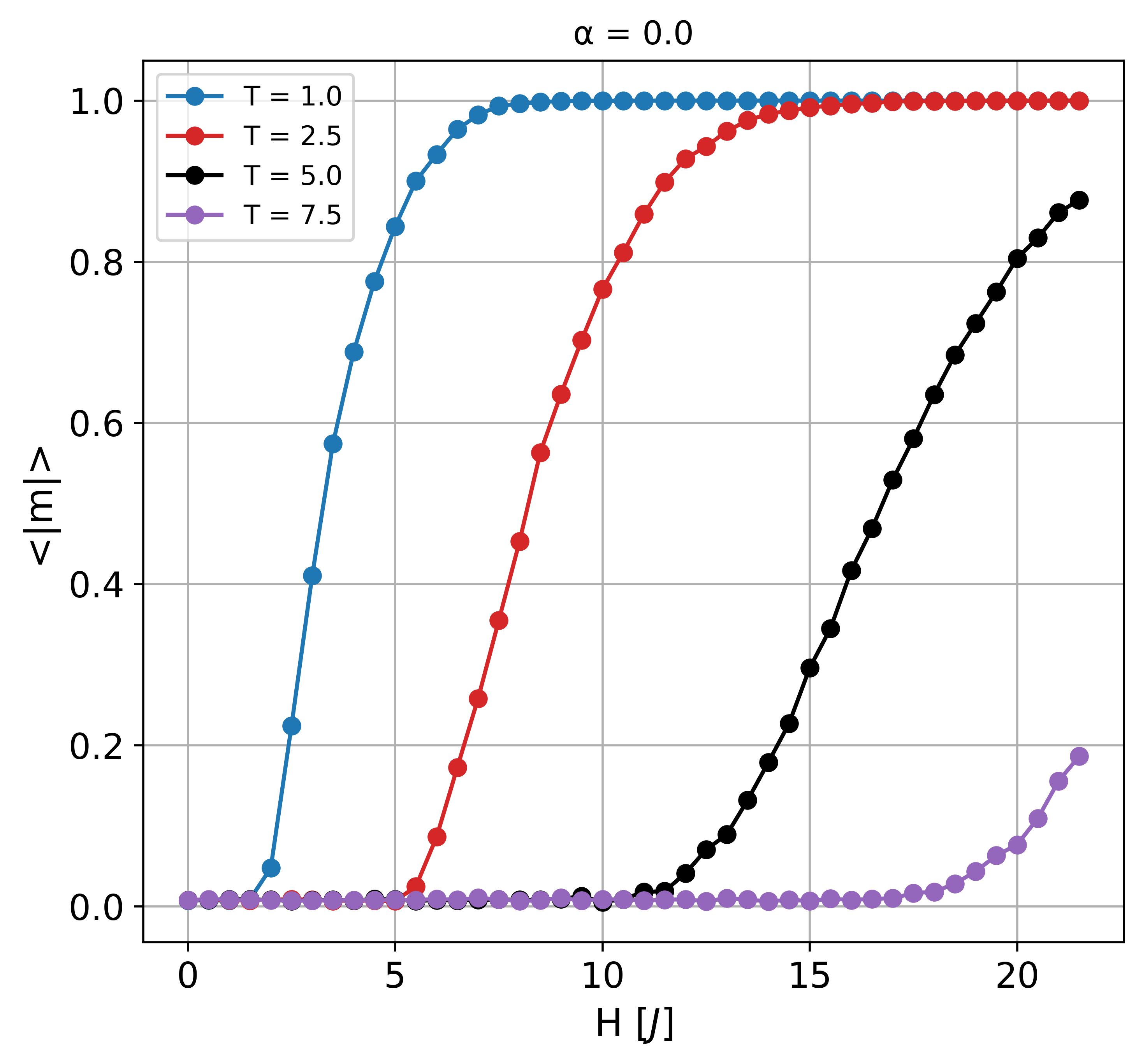}
    \caption{Field dependence of the average magnetization per site for $\alpha = 0$ and several temperatures. The system becomes more insensitive to respond to the field the higher the temperature.}
    \label{fig:fig_1}
\end{figure}

\begin{figure}
    %\justifying
    \includegraphics[width=8cm,
    height=7cm]{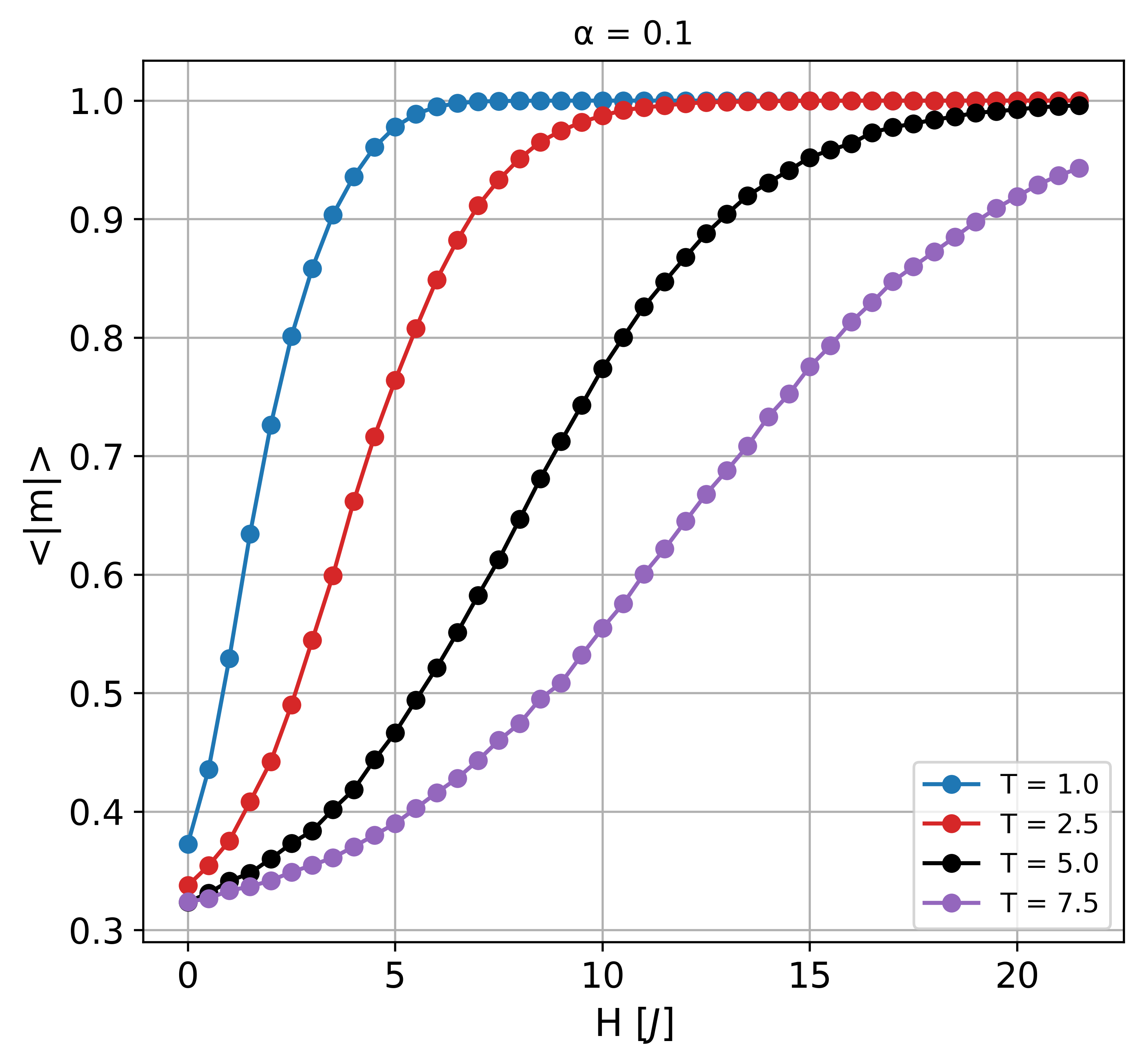}
    \caption{Field dependence of the average magnetization per site for $\alpha = 0.1$ and several temperatures. Differently from $\alpha = 0$, a low auditing probability makes the system to start responding quickly to field action, and saturation, which is in correspondence with a smaller tax evasion, is reached earlier at lower fields.}
    \label{fig:fig_2}
\end{figure}

\begin{figure}
    %\justifying
    \includegraphics[width=8cm,
    height=7cm]{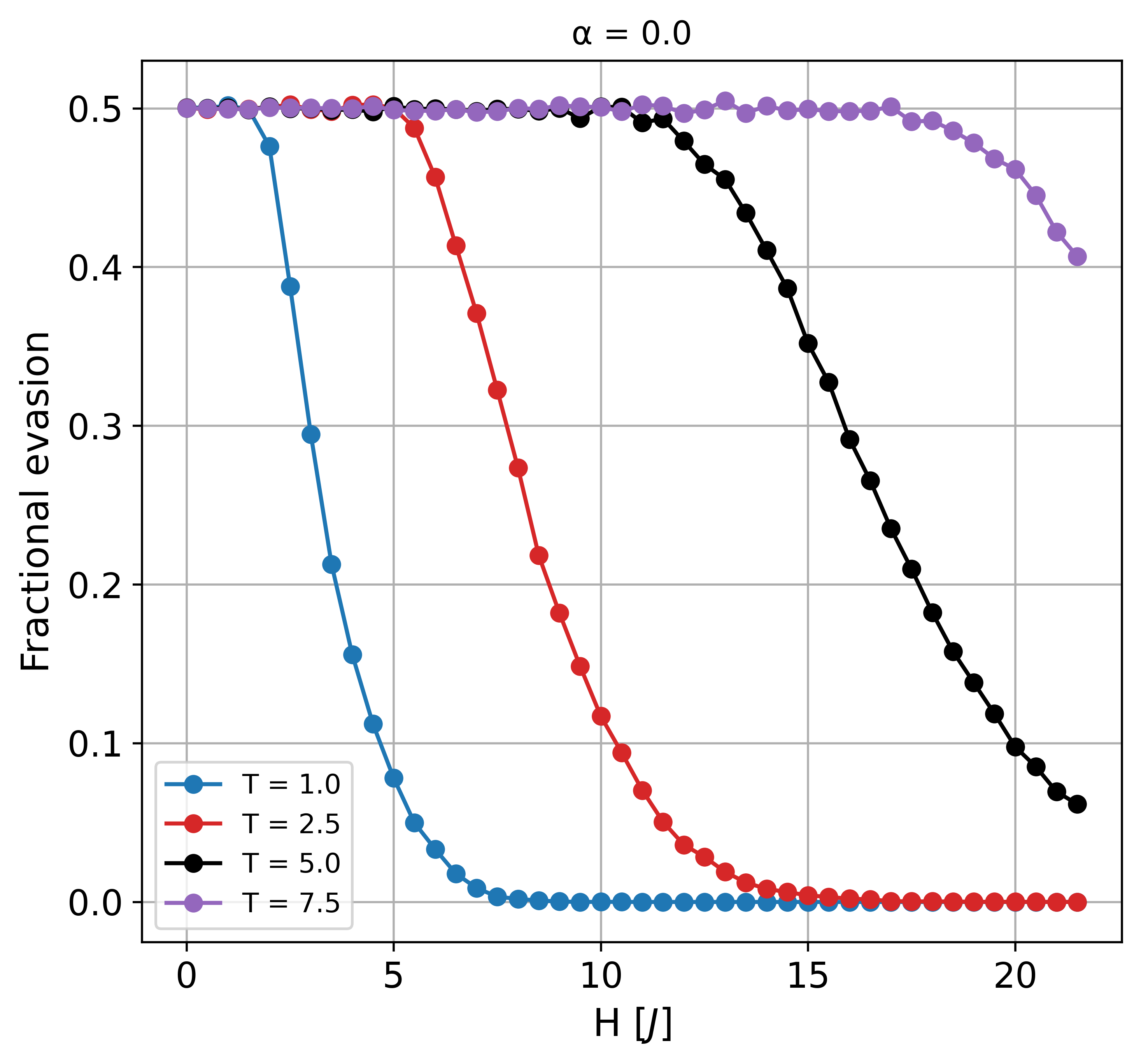}
    \caption{Field dependence of the average fractional evasion for $\alpha = 0$ and several temperatures. The system demands more restrictive policies, i.e. greater field values, to reduce evasion as the temperature increases.}
    \label{fig:fig_3}
\end{figure}

\begin{figure}
    %\justifying
    \includegraphics[width=8cm,
    height=7cm]{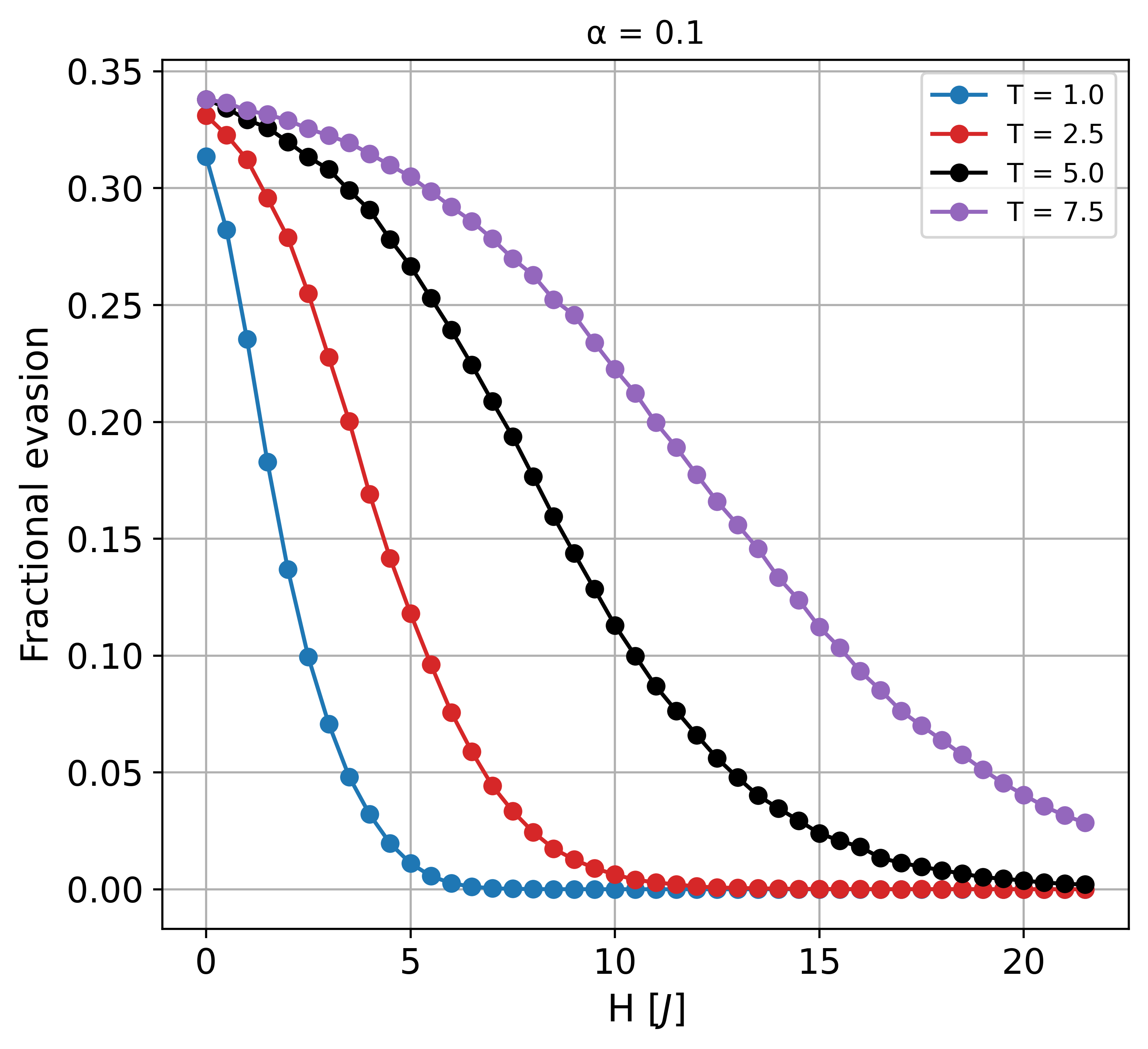}
    \caption{Field dependence of the average fractional evasion for $\alpha = 0.1$ and several temperatures. The system begins with a smaller zero-field fractional evasion than that for $\alpha = 0$, and a smaller tax evasion is reached earlier at lower fields or less restrictive policies.}
    \label{fig:fig_4}
\end{figure}

Due to the variety of parameters that have an effect on the system, first we start characterizing the degree of influence of the external field over the average magnetization per site, $\langle |m|\rangle$. Results are shown in figures \ref{fig:fig_1} and \ref{fig:fig_2} for some fixed temperatures and for audit probabilities $\alpha=0$ and $\alpha=0.1$ respectively. These results allow to tune both the range of temperature values to be used and the range of field values where the government policies, or equivalently, the external applied field, can saturate the system or to show a trend in such direction. This fact is important because the tendency to saturate the system corresponds to less evasion. Two important differences are noticed in these figures. For $\alpha=0$, the system begins to follow the field, at greater field values, the higher the temperature. Contrary to this, a low auditing probability, namely $\alpha=0.1$, makes the system to respond quickly to the field action as is observed in figure \ref{fig:fig_2}, and the saturation state is reached at smaller field values compared to those in figure \ref{fig:fig_1}. These characteristics, traduced in terms of the average fractional evasion are correspondingly shown in figures \ref{fig:fig_3} and \ref{fig:fig_4}.

As can be observed in figures \ref{fig:fig_3} and \ref{fig:fig_4}, in the framework of a policy free of audit ($\alpha = 0$), more restrictive global or nation policies or even a higher degree of favorability of the perception of society towards the government of the day, which implies a greater $H$ value, must be applied to reduce evasion, but if the complexity of such policies or the strength of $H$ increases, more time (temperature) will be required to fulfill the policies. On the other hand, when agents are audited randomly with a non zero probability, e.g. $\alpha = 0.1$, the global government policies for tax collection can be less restrictive to get a significant reduction of evasion in a shorter time. This is what is observed in figure \ref{fig:fig_4}. In this way, tax collection becomes more effective if auditing policies are implemented in agents randomly chosen with a certain probability.

\subsection{Temperature dependence and phase transitions}

\begin{figure}
    \includegraphics[width=8cm,
    height=7cm]{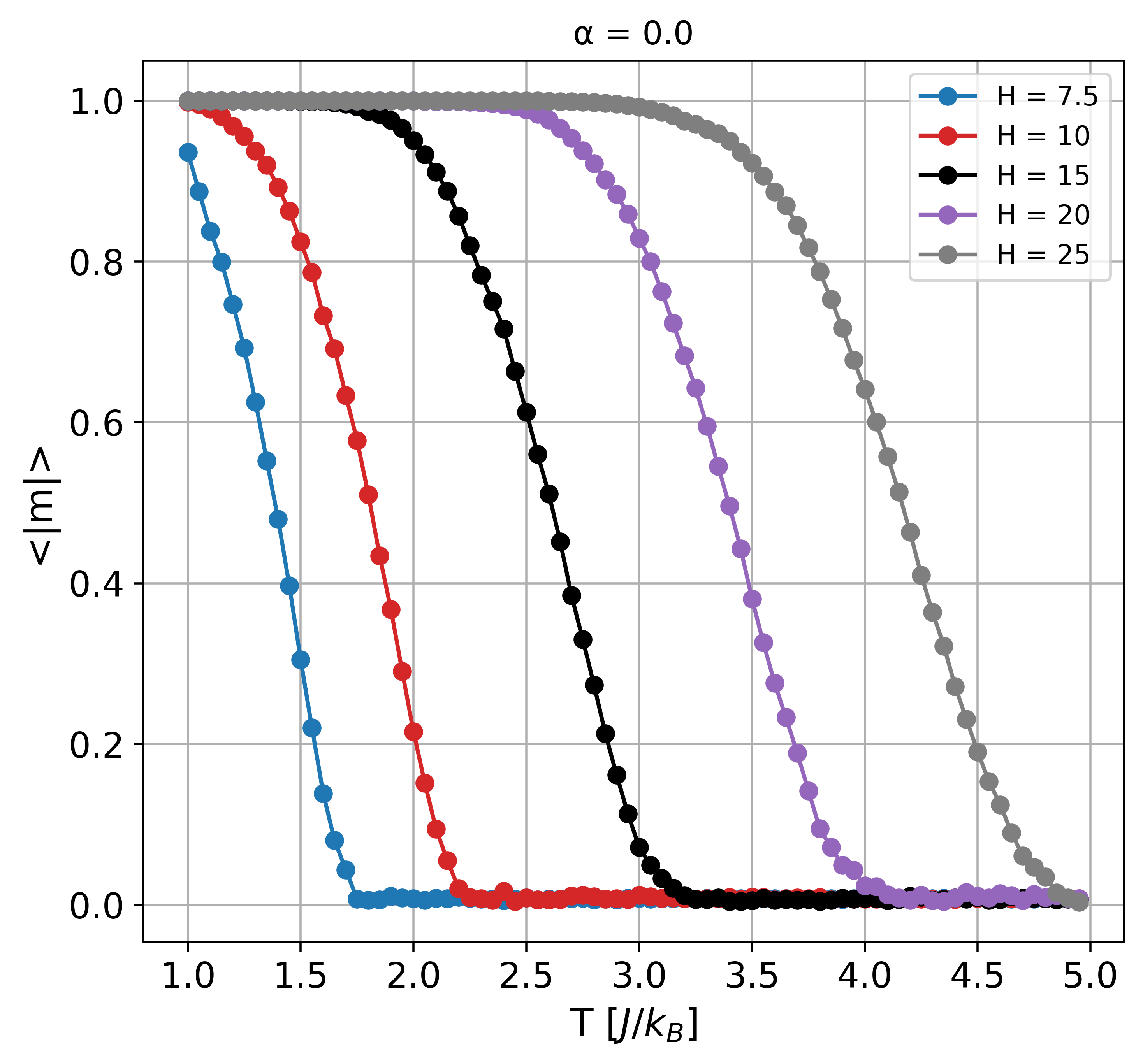}
    \caption{Temperature dependence of the average magnetization per site for $\alpha = 0$ and several field values. It can be noticed the shift of the inflection point towards greater values of $T$ as the strength of the external field increases.}
    \label{fig:fig_5}
\end{figure}

\begin{figure}
    \includegraphics[width=8cm,
    height=7cm]{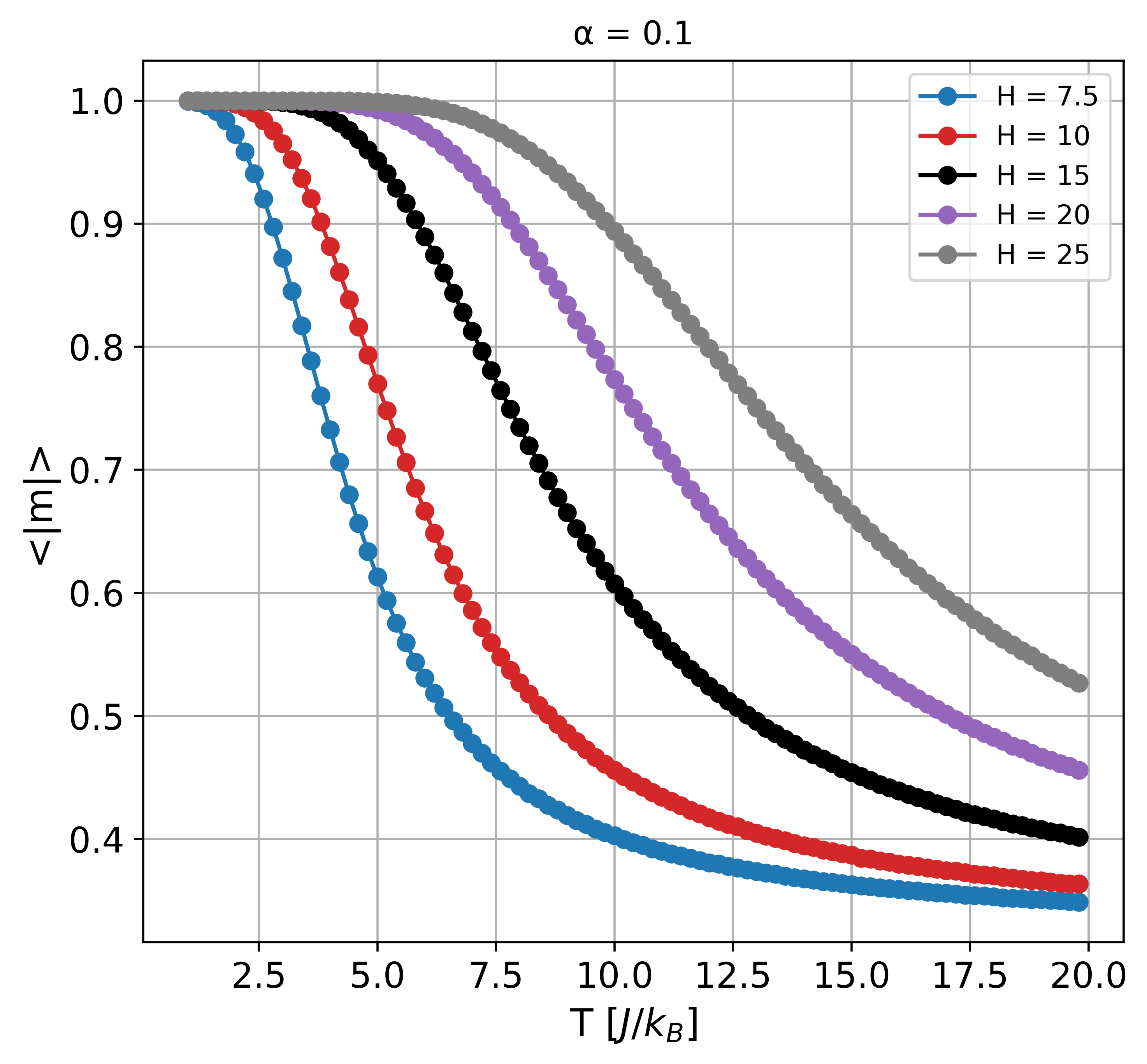}
    \caption{Temperature dependence of the average magnetization per site for $\alpha = 0.1$ and several field values. Here, the transition is delayed and shifted to higher $T$ values as observed in the $T$-range. The final values for $\langle |m|\rangle$ are greater than those of fig. \ref{fig:fig_5} above $T_c$. This effect is due to the third term in eq.\ref{eqn:eqn1} playing the role of a Zeeman interaction.} 
    \label{fig:fig_6}
\end{figure}

Since in the approach of this work, temperature is a global parameter that meets the characteristics of a thermal bath of a canonical ensemble, and it is related to the time needed for policies to be effectively applied, it is important to analyze the behavior of the observables as a function of temperature. Figures \ref{fig:fig_5} and \ref{fig:fig_6} describe the interplay between magnetic field and temperature for two different audit probabilities, namely $\alpha = 0$ and $\alpha = 0.1$, respectively. In both figures, the temperature dependence of the average magnetization per site is shown. The trend observed, reveals the typical behavior of a magnetic system undergoing a thermal-driven phase transition, from a ferromagnetic (FM) phase to a paramagnetic (PM) one. This is endorsed by the lambda-type behavior of the magnetic susceptibility discussed below. The temperature at which the peak of the susceptibility takes place, is related with an ordering or transition temperature $T_c$. This temperature is shifted, in both cases, towards higher values of $T$ as the strength of the external applied field $H$ increases and also if the audit probability $\alpha$ increases, since this parameter plays the role of a locally applied field with the same form of the Zeeman interaction.

\begin{figure}
    \includegraphics[width=8cm,
    height=7cm]{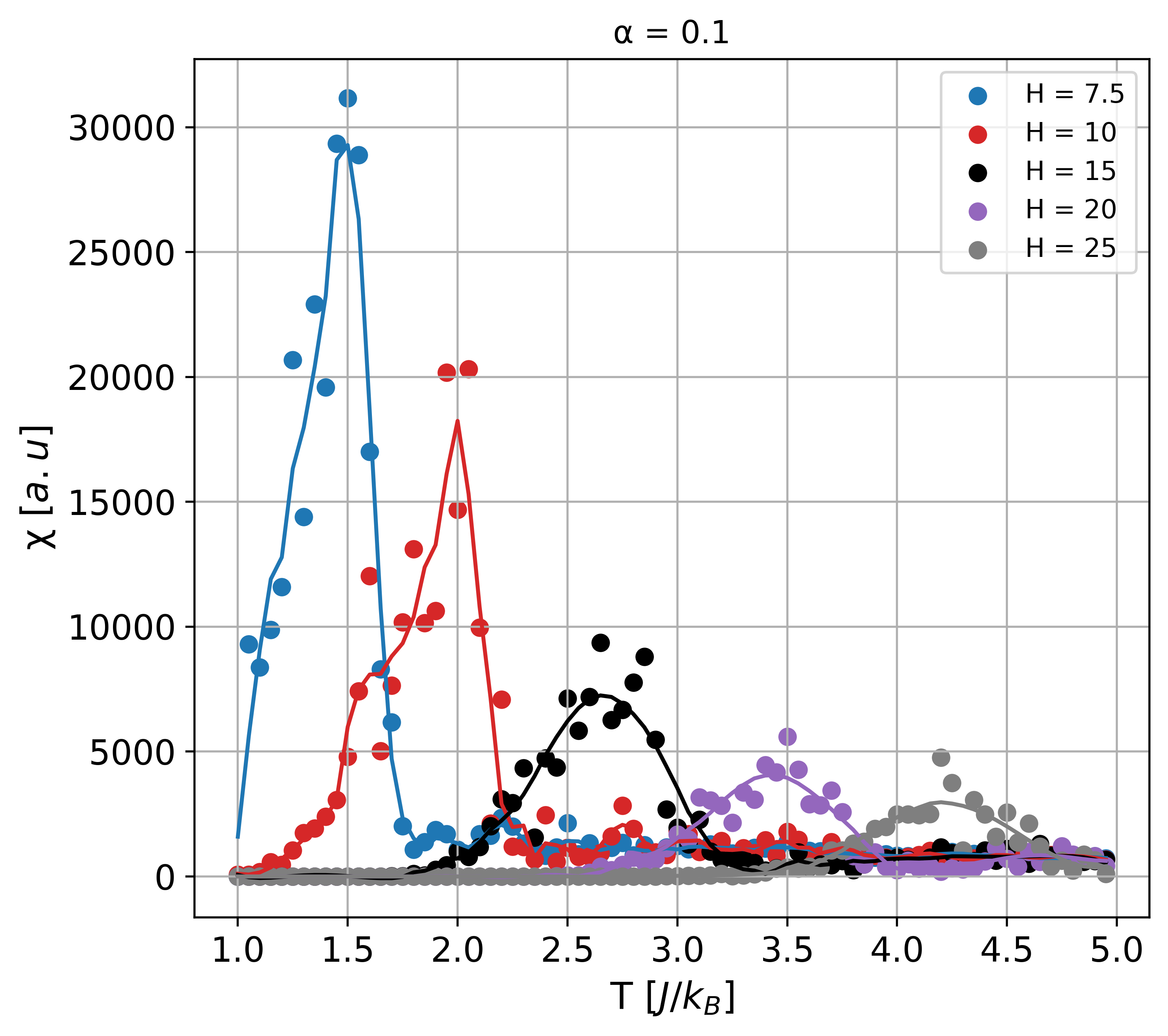}
    \caption{Temperature dependence of the magnetic susceptibility for $\alpha = 0$ and several field values. The location of the peaks correspond to $T_c$ and they are shifted to higher $T$ values as $H$ increases. Lines are guides to the eye.}
    \label{fig:fig_7}
\end{figure}

\begin{figure}
    \includegraphics[width=8cm,
    height=7cm]{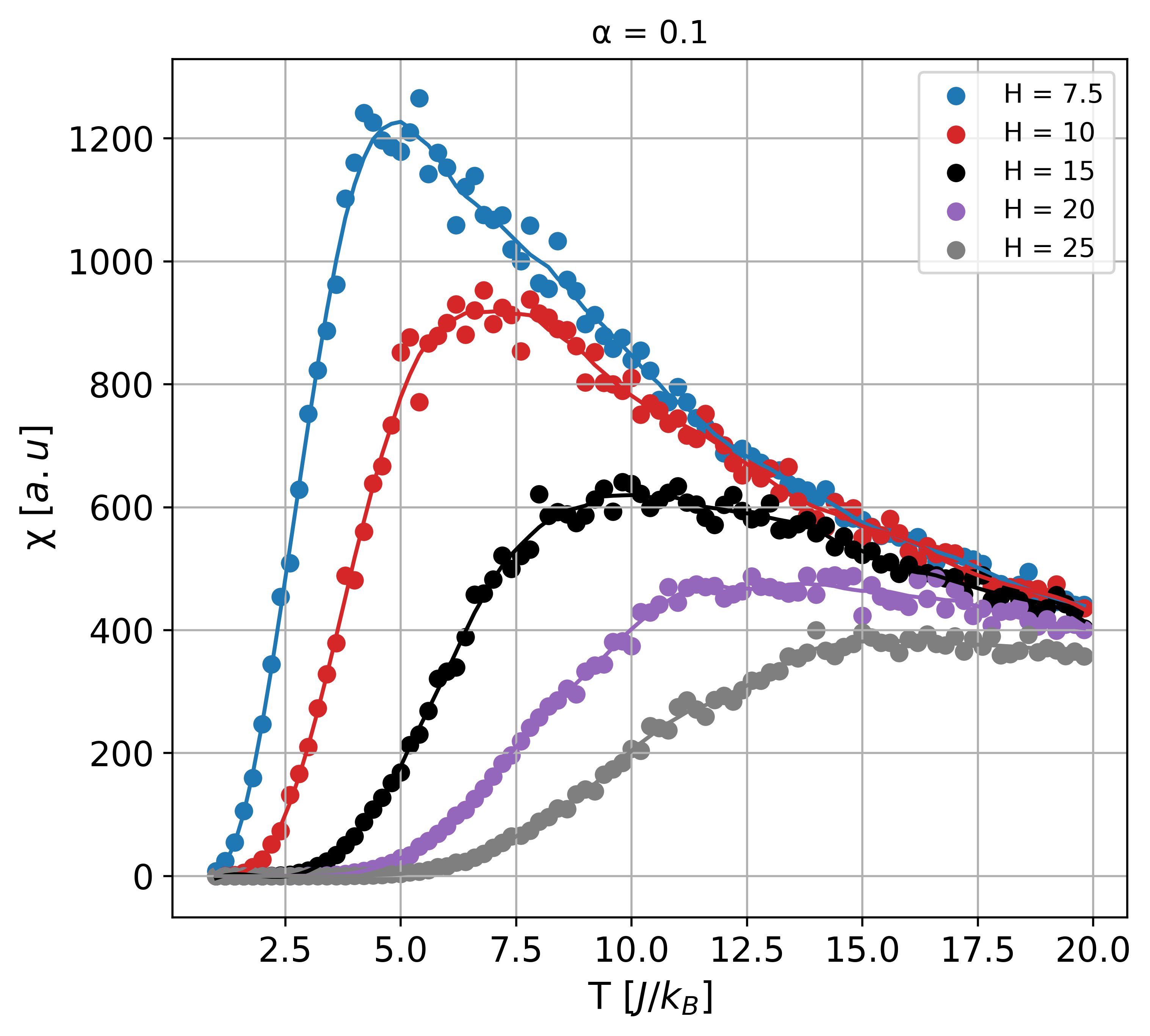}
    \caption{Temperature dependence of the magnetic susceptibility for $\alpha = 0.1$ and several field values. The location of the peaks correspond to $T_c$ and they are shifted to higher $T$ values as $H$ increases and they become more rounded with respect to the figure \ref{fig:fig_7}. Lines are guides to the eye.} 
    \label{fig:fig_8}
\end{figure}

The difference between having or not an audit probability, lies in the fact, that a sharp transition from an ordered state to a disordered one is observed in the vicinity of $T_c$ for $\alpha = 0$, whereas for $\alpha = 0.1$ such transition is more gradual and more extended along the temperature line. This difference is reflected in the degree of rounding, width and height of the peak of the magnetic susceptibility, given by $\chi = (1/k_{B}T)(\langle m^{2} \rangle - \langle |m| \rangle^{2})$, which is shown in figures \ref{fig:fig_7} and \ref{fig:fig_8} for the two $\alpha$ values considered. On the other hand the shift of the peak, from a social perspective, can be interpreted as an indicative of the need to require longer implementation and compliance times as stronger and more elaborate government policies are included like those derived, for instance, from a tax reform. This fact is reinforced as auditing tasks are involved, where the transition becomes more gradual and continuous.

\begin{figure}
    \includegraphics[width=8cm,
    height=7cm]{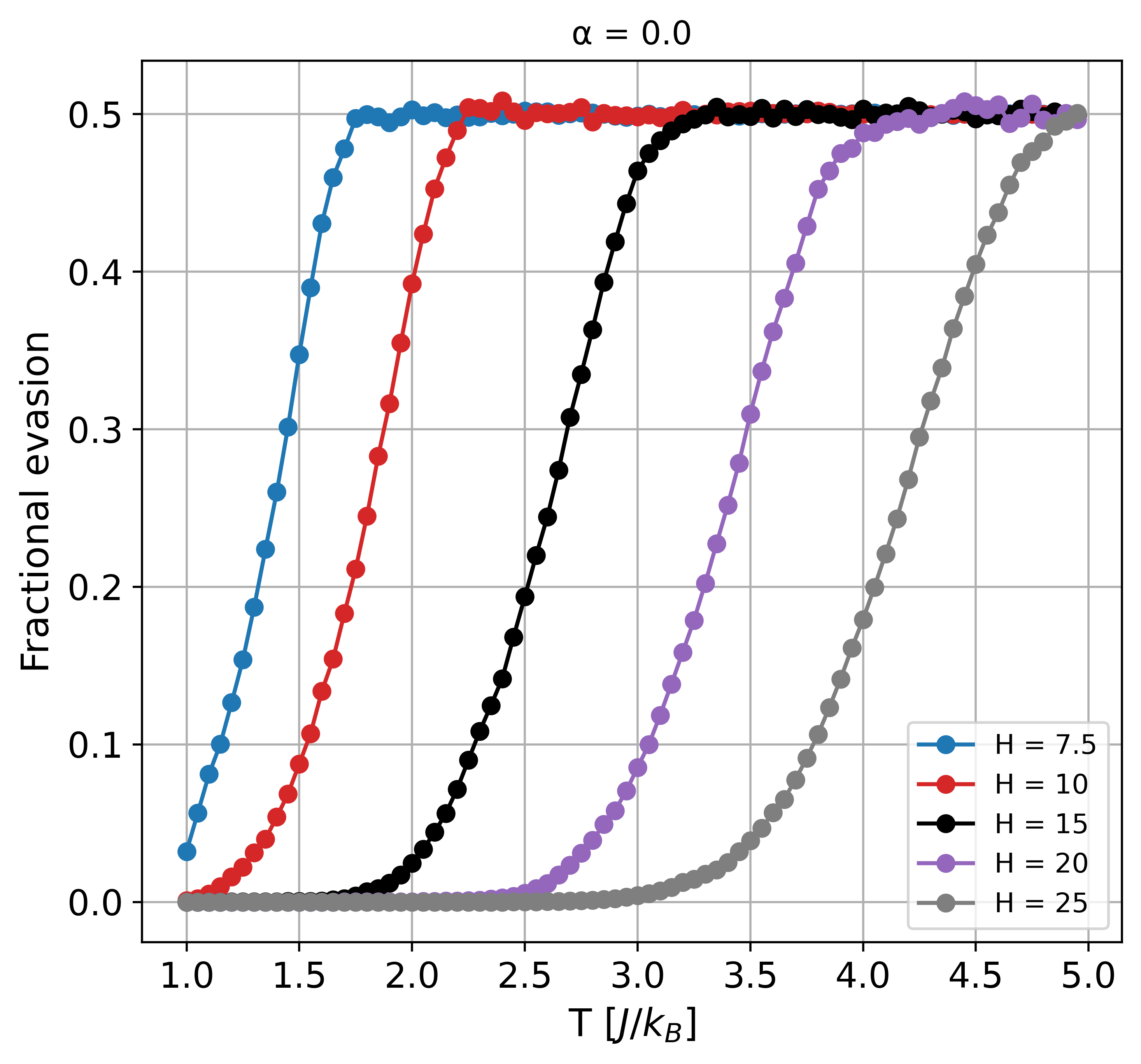}
    \caption{Temperature dependence of the average fractional evasion for $\alpha = 0$ and several field values.} 
    \label{fig:fig_9}
\end{figure}

\begin{figure}
    \includegraphics[width=8cm,
    height=7cm]{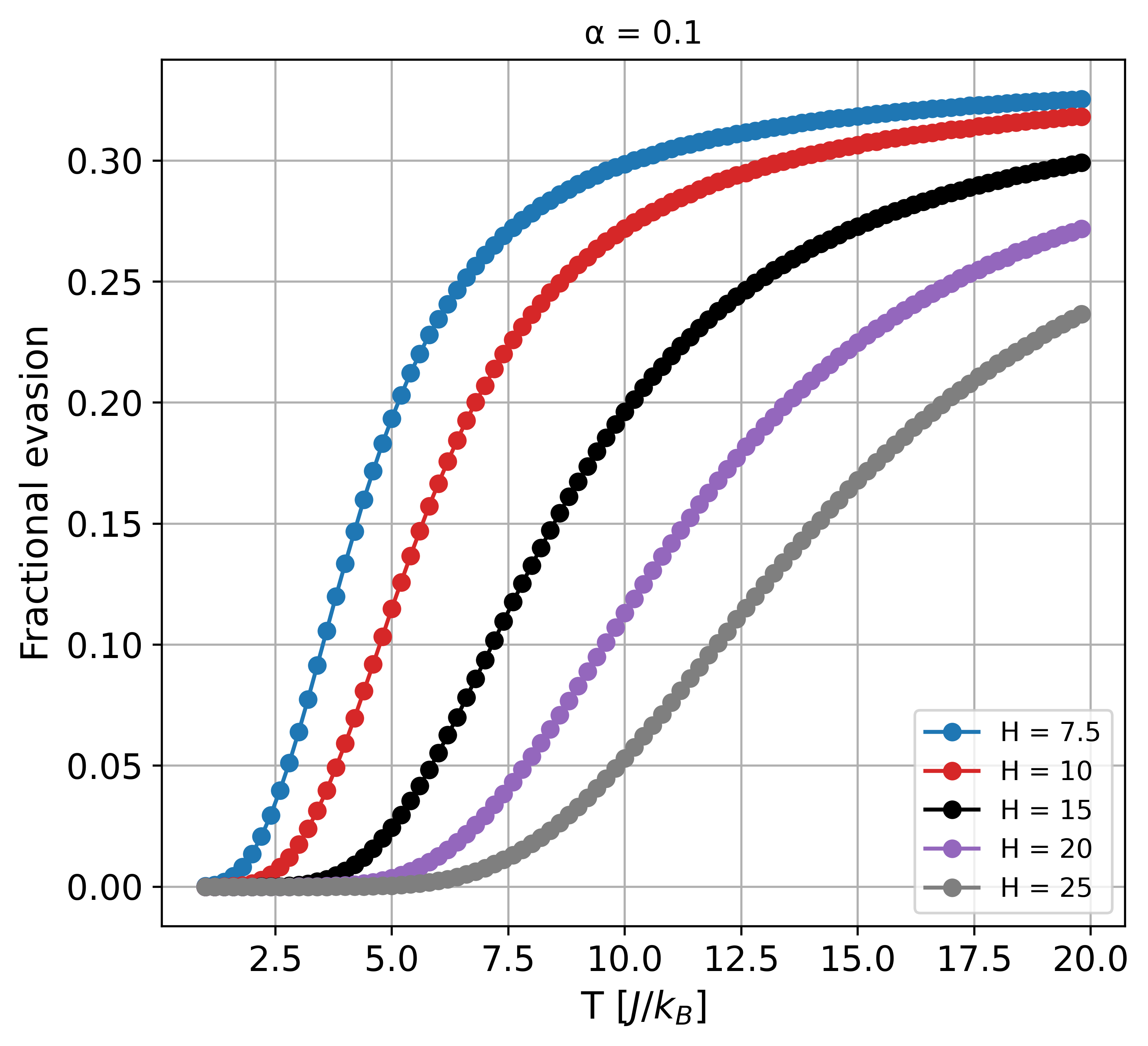}
    \caption{Temperature dependence of the average fractional evasion for $\alpha = 0.1$ and several field values.} 
    \label{fig:fig_10}
\end{figure}

 To see how the behavior of the magnetization is reflected in terms of evasion, the corresponding results are shown in figures \ref{fig:fig_9} and \ref{fig:fig_10}. As is noticed, the FM phase below $T_c$ is in correspondence with a compliant taxpayers state linked to a low evasion, whereas the PM phase is consistent with a state of agents not paying their tribute associated to a high evasion.
 
As we can infer from figures \ref{fig:fig_9} and \ref{fig:fig_10}, a fractional evasion of $50$\% is reached in the absence of any audit. In addition, evasion begins to manifest rapidly in the low-temperature regime from some characteristic time ($T_c$), which is dependent on the degree of severity ($H$) of policies implemented by the government. The tougher the policies, the longer it takes for society to evade, without this representing a decrease in maximum avoidance. Differently from this, if an audit scheme is implemented, e.g. with $10$\% of probability, i.e. $\alpha = 0.1$, evasion does not disappear, but it is strongly reduced as can be observed in the maximum values for evasion in figure \ref{fig:fig_10}, which are lower than those of figure \ref{fig:fig_9} for $\alpha = 0$. This means when there is not penalty, the levels of evasion are higher. Results also show that for a given time (or temperature) evasion is reduced with a tightening of collection policies. Besides, according to the greater temperature range used, more time is needed to achieve evasion saturation, which is in any case less than that for $\alpha = 0$. 

\subsection{Effect of the audit probability on tax evasion}

\begin{figure}
    \includegraphics[width=8cm,
    height=7cm]{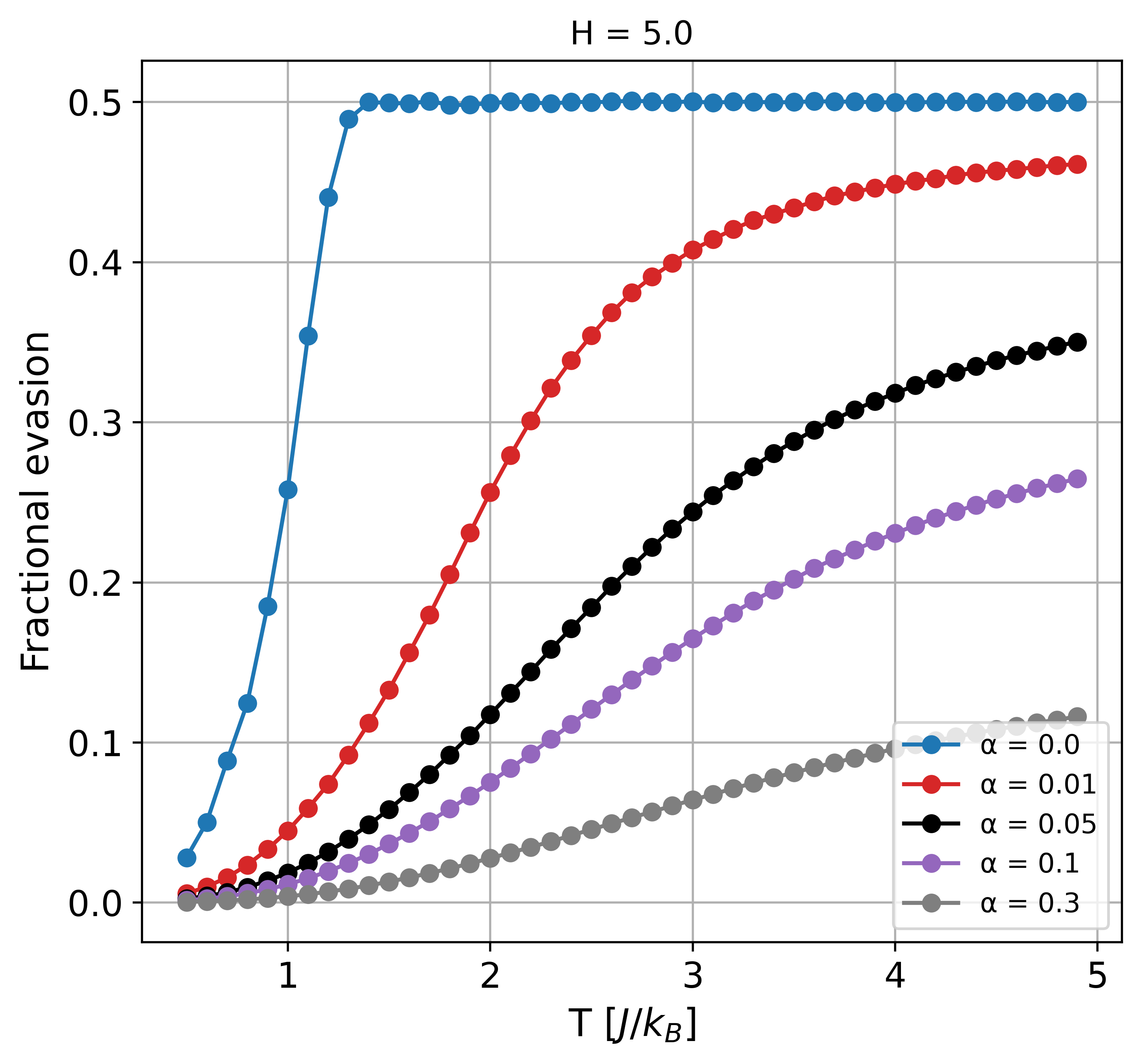}
    \caption{Temperature dependence of the average fractional evasion for $H = 5.0$ and several $\alpha$ values.} 
    \label{fig:fig_11}
\end{figure}

A more detailed analysis of the effect of the audit probability on the evasion for a given field value relatively low, for instance $H=5.0$, which means general soft government policies, is shown in figure \ref{fig:fig_11}. From here, it is clear that a greater reduction in evasion is achieved as the audit probability is greater. Moreover, even with a very low audit probability, the differences respect to the case $\alpha = 0$, are remarkable. This means that large audit probability values are not needed to obtain an effective control over tax evasion.

\begin{figure}
    \includegraphics[width=8cm,
    height=7cm]{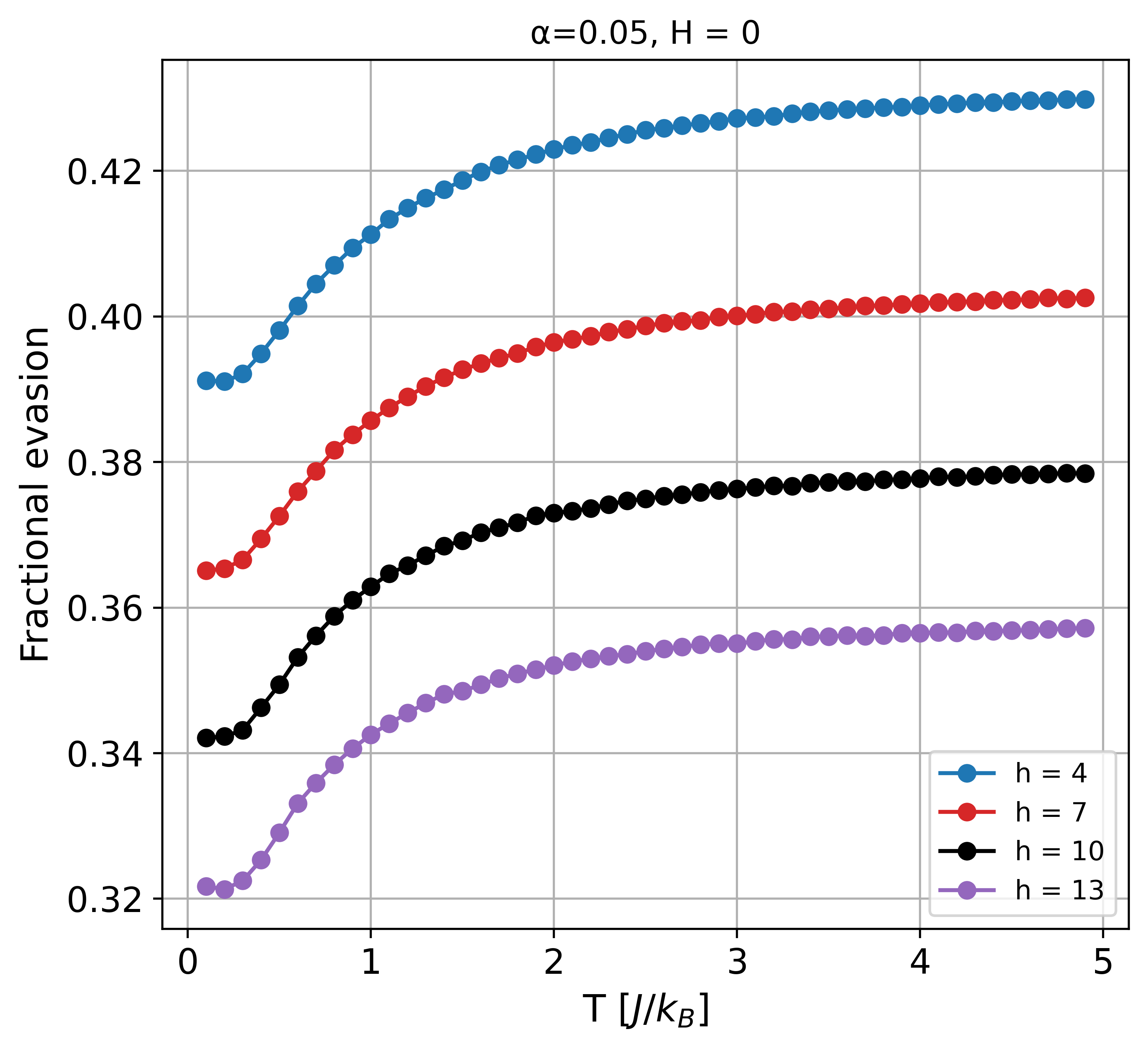}
    \caption{Temperature dependence of the average fractional evasion for $H=0$ and $\alpha = 0.05$, and for several values of the the penalty rate $h$.}
    \label{fig:fig_12}
\end{figure}

Regarding the parameter $h$ in eq.\ref{eqn:eqn1}, this refers to the number of fiscal periods under which the policies of a given country can enforce to pay the respective tax debt to an evader agent randomly chosen with an audit probability $\alpha$, or at least to exert greater control and surveillance during a period of time. In the results presented so far, we have randomly chosen this value to be within the range $h \in [4,...,9]$, consistent to the ones considered in other works ~\cite{Zaklan2009,Seibold2013b,Crokidakis2014a}. To analyse the individual effect of $h$ over evasion, we can consider in principle a null governance system ($H = 0$) with a relatively low audit probability over the population of agents (e.g. $\alpha = 0.05$ or $5$\%), but where the key parameter is the time dedicated to exert control over those evaders randomly chosen. The corresponding results are shown in figure \ref{fig:fig_12}. When the audit probability is very low ($\alpha = 0.05$) and $H = 0$, the system in principle is supposed to be free enough to easily decide between payment or evasion. However this assertion is limited by the number of periods under which an evader will be submitted to control and surveillance in order to remain honest. Higher the penalization via $h$, smaller the evasion. By comparing with the results of figure \ref{fig:fig_11}, high penalty rates can be replaced therefore by the presence of a government with soft public policies in general terms, but with a strong presence of local auditing along the time.

\begin{figure}
    \includegraphics[width=8cm,
    height=7cm]{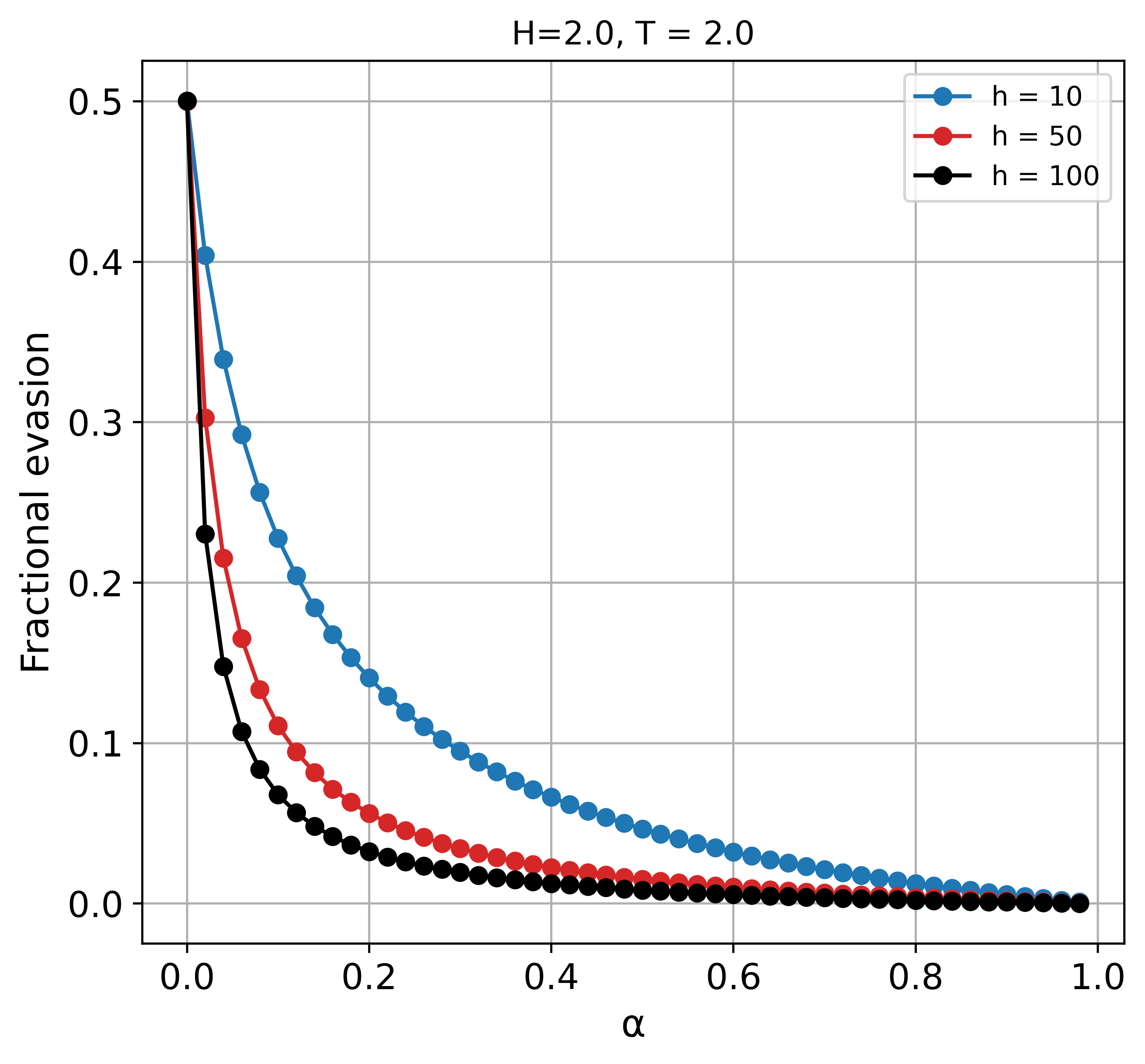}
    \caption{Average fractional tax evasion as a function of the audit probability $\alpha$ for $T=2 J/k_B$, $H=2.0$ and for the same $h$ values used in reference \cite{Crokidakis2014a}.}
    \label{fig:fig_13}
\end{figure}

Another interesting perspective is the dependence of the average fractional evasion with the audit probability at a given temperature $T$ and several penalization periods shown in figure \ref{fig:fig_13}. As can be observed, evasion decreases in a hyperbolic fashion as the audit probability increases, and such behavior is more pronounced as the number of penalization periods becomes greater. This endorses the fact that auditing is an effective way of diminishing the levels of evasion in a society along with the intensity of the surveillance period $h$. These results are in agreement with those obtained from the three-state kinetic agent-based model of reference \cite{Crokidakis2014a} (see fig. 5(b) in that work), where average stationary tax evasion as a function of the audit probability was obtained. The number of penalization periods in that work was labeled as $k$ and corresponds to the number of periods that a detected tax evader must remain honest like in our case.

\subsection{Effect of exchange integrals}

\begin{figure}
    \includegraphics[width=8cm,
    height=7cm]{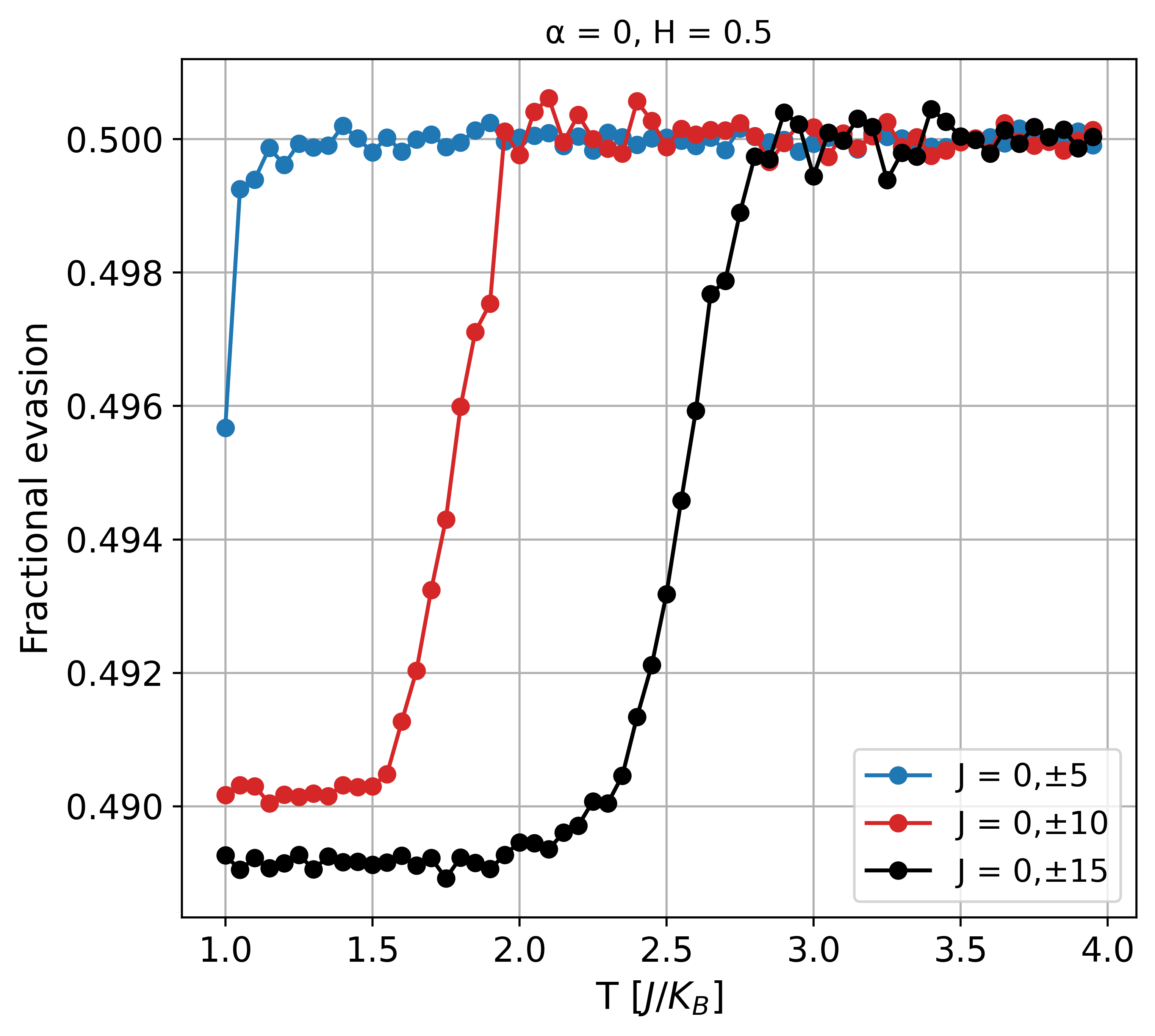}
    \caption{Temperature dependence of the average fractional evasion for $\alpha = 0$ and $H=0.5$. For no audit probability, and low influence of the government policy, exchange integrals $J_{ij}$ only play a role in the transition region towards the stationary state, no matter how ``strong" the exchange of information is.}
    \label{fig:fig_14}
\end{figure}

\begin{figure}
    \includegraphics[width=8cm,
    height=7cm]{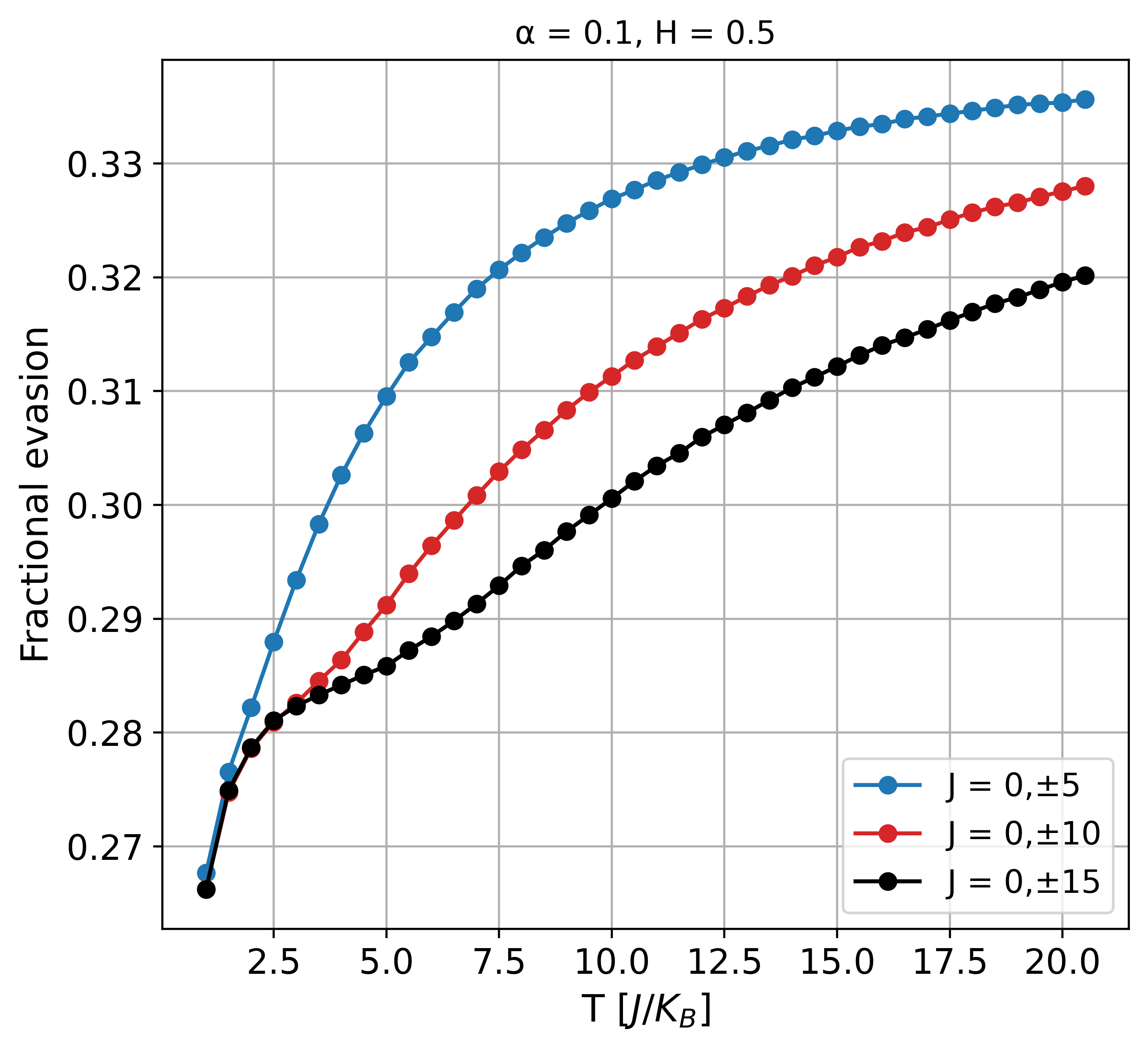}
    \caption{Temperature dependence of the average fractional evasion for $\alpha = 0.1$ and $H=0.5$.}
    \label{fig:fig_15}
\end{figure}

In this section we study how the values of the exchange integrals $J_{ij}$ in eq. \ref{eqn:eqn1} can affect tax evasion. These integrals play the social role of the information that an agent can receive from the immediate or local surroundings or nearest neighbors. For this purpose, simulation was performed for audit probability values $\alpha = 0$ and $\alpha = 0.1$. Figure \ref{fig:fig_14} shows the average fractional evasion as a function of temperature for different sets of exchange integrals for $\alpha = 0$ and $H=0.5$. As is observed, the values taken by $J_{ij}$ are only important to determine the path through which the system reaches the steady state, but they do not affect the final evasion percentage once thermalization has been reached. This means that, the influence due to nearest neighbors over the agents does not play a crucial role in terms of final evasion for a null audit probability and a low intensity of the nation regulatory policy $H$. A similar behavior is attained, as is noticed in figure \ref{fig:fig_15}, for $\alpha = 0.1$ and $H=0.5$. In this last case, evasion fractions are smaller than the previous ones as a consequence of the audit probability and the stationary state is achieved at higher temperatures. The short range information provided by $J_{ij}$ helps keeping undecided the agents for longer periods, favoring at the end the reduction of tax evasion near the stationary state. Additionally, different from the case with $\alpha = 0$, the steady state, i.e. the final value for evasion, is now dependent on the value of $J_{ij}$.

\subsection{Colombia as a case study}

\begin{table*}[ht]
\begin{center}
\begin{tabular}{|l|l|l|l|l|l|l|l|l|l|l|l|l|l|l|l|l|l|}
\hline
 Year & 2000 & 2001 & 2002 & 2003 & 2004 & 2005 & 2006 & 2007 & 2008 & 2009 & 2010 & 2011 & 2012 & 2013 & 2014 & 2015 & 2016 \\ \hline
 Percentage & $48.0$ & $46.7$ & $48.1$ & $41.8$ & $29.9$ & $37.9$ & $46.5$ & $40.5$ & $37.8$ & $45.9$ & $43.0$ & $34.0$ & $33.6$ & $34.9$ & $38.7$ & $38.2$ & $36.3$  \\ \hline
\end{tabular}
\caption{Tax evasion percentages for Colombia over a period of 17 years \cite{Cuervo2018}.}
\label{tab:tab_1}
\end{center}
\end{table*}

\begin{table*}[ht]
\begin{center}
\begin{tabular}{|l|l|l|l|}
\hline
\multicolumn{4}{|l|}{$\alpha$ and $H$ values to obtain $\bar{f.ev}$}\\
\hline \hline
Year  & $H$ & $\alpha$ & $Evasion$ \\ \hline
2001 & $0.020$ & $0.015$ & 46.9\% \\ \hline
2004 & $0.720$ & $0.130$ & 29.8\%  \\ \hline
2007 & $0.014$ & $0.051$ & 40.6\%  \\ \hline
2012 & $0.290$ & $0.100$ & 33.6\%  \\ \hline
\end{tabular}
\caption{$\alpha$ and $H$ values used to fit some of the fractional evasion results published by Rodriguez-Cuervo \cite{Cuervo2018}. It can be seen that a nation government policy in conjunction with and audit probability should be articulated to obtain an effective tax evasion reduction as seen in the 2004 results.}
\label{tab:tab_2}
\end{center}
\end{table*}

\begin{figure}
    \includegraphics[width=8cm,
    height=7cm]{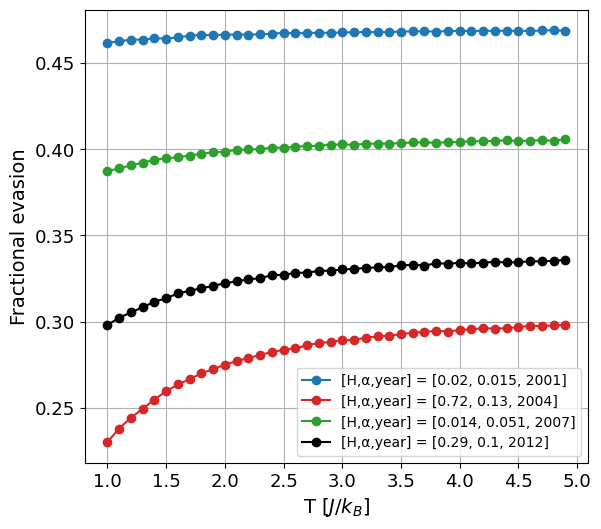}
    \caption{Average fractional evasion as a function of $T$ for some selected years and by using the fitting parameters $(H, \alpha)$ presented in table \ref{tab:tab_2}.}
    \label{fig:fig_16}
\end{figure}

Finally, we use our model to fit the tax evasion percentages reported by Rodriguez-Cuervo ~\cite{Cuervo2018} for the Republic of Colombia. In that work, an analysis of the tax evasion in Colombia during the period 2000-2016 is presented according to the statistics of the National Tax and Customs Directorate (DIAN). The respective data are summarized in table \ref{tab:tab_1}. An interesting feature from these data is the lowest evasion percentage ($29.9\%$) recorded in 2004, presumably due to the fact that, in August 2002, the government of that time imposed, as a measure to finance the armed conflict in Colombia, the so-called wealth tax to ``cover the expenses necessary to meet democratic security", determining that the taxpayers of the tax would be all income taxpayers, thus expanding the collection base. This historical milestone, in the fight to achieve the peace in Colombia, implied both a tax reform and a strong control and vigilance with the collection. This fact could be interpreted, in the light of our model, to an increase in both $H$ and $\alpha$ parameters. On this basis, the values for $H$ and $\alpha$ in our model were fitted in order to reproduce the tax evasion percentages in that period of time. However, it must be stressed that different sets of pairs ($H$,$\alpha$) can in principle be solution for a given tax evasion, making the system multiply degenerated, which is a main characteristic of the Ising-type spin-glass systems. The corresponding results are shown in fig. \ref{fig:fig_16} and table \ref{tab:tab_2} summarizes the parameters used to fit experimental data. As observed, a nation government policy $H$ in conjunction with and audit probability $\alpha$ should be articulated to obtain an effective tax evasion reduction as seen in the 2004 results, where the highest $H$ and $\alpha$ values were obtained.

Therefore, our model can be used as a tool to project the relationship between nation tax policies and the audit probability. For instance, between $2001$ and $2004$, $\Delta \alpha = 0.115$, and $|\Delta H| = 0.70$, while for the period $2007-2012$, $\Delta \alpha = 0.049$, and $|\Delta H| = 0.276$. This fact implies that, even when the image of the government can remain positive or favorable, it is not a guarantee that there will be a reduction in evasion. A relative strong or popular government, together with a confuse or weak tax policy can barely lead to some reduction of the evasion (as the 2012 case in table \ref{tab:tab_2}). On the other hand, in 2004 not only the government had a good interaction with the agents though a well defined tax reform, but the audit probability increased and the tax policy improved, leading to $\%ev = 29.9\%$, the lowest value found in the period studied in ~\cite{Cuervo2018}. 

\section{Conclusions}
\label{Conclusions_Sec}

The above results reveal that, from the perspective of the model stated here, the eq.\ref{eqn:eqn1} under the conditions of eq.\ref{eqn:eqn3} provides good results, which cover, not only works done before in the topic of tax evasion from an econophysics approach like the three-state kinetic agent-based model in \cite{Crokidakis2014a}, but also, allows to replicate results obtained by economical means as we have shown for the Colombian case. Even more, the model allows to identify the interaction between government policy and audit probability in order to reduce tax evasion. Under the perspective of this model, magnetic field and $\alpha$, or equivalently, government and local audit, must be complementary if a reduction in tax evasion is desired. Recalling the values of table \ref{tab:tab_2}, an effective tax policy is reached by a joint action of a reliable government in which the agents can trust, and a strict tax policy. Although in practice can be impossible to audit the $100\%$ of population, this can be optimized with a suitable relationship between agents - government policies and society in general.

From a physical perspective, we have identified the system composed by agents susceptible of tax payment with a ternary disordered and diluted alloy of the form $A_{p}B_{x}C_q$, with an Ising-type diluted system  with competing interactions. Such competing interactions and the degree of dilution can model the personality of the agents when making a decision, and safeguard the temperature as a global or bath parameter, differently from other authors where temperature has been regarded as a changing on-site local parameter. Even though we have proposed that temperature is related to time, its real-time equivalence is still an open issue and it is is going to depend on the country's legal framework.

The number of periods $h$ or penalty rates also play an important role for reducing evasion, provided there is a high degree of accompaniment, control and surveillance. On the other hand, regarding the influence of the exchange integrals $J_{ij}$ with and without audit probability, we can conclude that, in the absence of audit, the influence of sharing information does not modify the stationary final evasion, whereas if an audit probability is present, sharing local information can modify effectively the decision made by the agents, with the consequent change in final evasion.

\acknowledgments
J.R acknowledges University of Antioquia for the exclusive dedication program and the CODI-UdeA 2017-16253 project. J.G-B acknowledges Biophysics of Tropical Diseases Max Planck Tandem Group for financial support and useful discussion. J.G-B and J.R acknowledge to Prof. Dr. Götz Seibold for providing us with the FORTRAN code used in \cite{Pickhardt2014a}, and for making the article referenced here as \cite{Berger2020} available upon request. All authors contributed equally to the work

\section{Additional information}

The authors declare no competing interests. Materials and data are promptly available for all readers.

\bibliography{Tax_Evasion}

\begin{thebibliography}{29}%
\makeatletter
\providecommand \@ifxundefined [1]{%
 \@ifx{#1\undefined}
}%
\providecommand \@ifnum [1]{%
 \ifnum #1\expandafter \@firstoftwo
 \else \expandafter \@secondoftwo
 \fi
}%
\providecommand \@ifx [1]{%
 \ifx #1\expandafter \@firstoftwo
 \else \expandafter \@secondoftwo
 \fi
}%
\providecommand \natexlab [1]{#1}%
\providecommand \enquote  [1]{``#1''}%
\providecommand \bibnamefont  [1]{#1}%
\providecommand \bibfnamefont [1]{#1}%
\providecommand \citenamefont [1]{#1}%
\providecommand \href@noop [0]{\@secondoftwo}%
\providecommand \href [0]{\begingroup \@sanitize@url \@href}%
\providecommand \@href[1]{\@@startlink{#1}\@@href}%
\providecommand \@@href[1]{\endgroup#1\@@endlink}%
\providecommand \@sanitize@url [0]{\catcode `\\12\catcode `\$12\catcode
  `\&12\catcode `\#12\catcode `\^12\catcode `\_12\catcode `\%12\relax}%
\providecommand \@@startlink[1]{}%
\providecommand \@@endlink[0]{}%
\providecommand \url  [0]{\begingroup\@sanitize@url \@url }%
\providecommand \@url [1]{\endgroup\@href {#1}{\urlprefix }}%
\providecommand \urlprefix  [0]{URL }%
\providecommand \Eprint [0]{\href }%
\providecommand \doibase [0]{https://doi.org/}%
\providecommand \selectlanguage [0]{\@gobble}%
\providecommand \bibinfo  [0]{\@secondoftwo}%
\providecommand \bibfield  [0]{\@secondoftwo}%
\providecommand \translation [1]{[#1]}%
\providecommand \BibitemOpen [0]{}%
\providecommand \bibitemStop [0]{}%
\providecommand \bibitemNoStop [0]{.\EOS\space}%
\providecommand \EOS [0]{\spacefactor3000\relax}%
\providecommand \BibitemShut  [1]{\csname bibitem#1\endcsname}%
\let\auto@bib@innerbib\@empty
%</preamble>
\bibitem [{\citenamefont {Teymur}\ and\ \citenamefont
  {Saman}(2012)}]{Teymur2012}%
  \BibitemOpen
  \bibfield  {author} {\bibinfo {author} {\bibfnamefont {R.}~\bibnamefont
  {Teymur}}\ and\ \bibinfo {author} {\bibfnamefont {F.}~\bibnamefont {Saman}},\
  }\bibfield  {title} {\bibinfo {title} {{Corruption, democracy and tax
  compliance: Cross-country evidence}},\ }\href@noop {} {\bibfield  {journal}
  {\bibinfo  {journal} {Bus. Manag. Rev.}\ } (\bibinfo {year}
  {2012})}\BibitemShut {NoStop}%
\bibitem [{\citenamefont {Lunina}\ \emph {et~al.}(2020)\citenamefont {Lunina},
  \citenamefont {Bilousova},\ and\ \citenamefont {Frolova}}]{Lunina2020}%
  \BibitemOpen
  \bibfield  {author} {\bibinfo {author} {\bibfnamefont {I.}~\bibnamefont
  {Lunina}}, \bibinfo {author} {\bibfnamefont {O.}~\bibnamefont {Bilousova}},\
  and\ \bibinfo {author} {\bibfnamefont {N.}~\bibnamefont {Frolova}},\
  }\bibfield  {title} {\bibinfo {title} {{Tax Reforms for the Development of
  Fiscal Space}},\ }\href {https://doi.org/10.30525/2256-0742/2020-6-3-48-58}
  {\bibfield  {journal} {\bibinfo  {journal} {Balt. J. Econ. Stud.}\ }\textbf
  {\bibinfo {volume} {6}},\ \bibinfo {pages} {48} (\bibinfo {year}
  {2020})}\BibitemShut {NoStop}%
\bibitem [{\citenamefont {Allingham}\ and\ \citenamefont
  {Sandmo}(1972)}]{Allingham1972}%
  \BibitemOpen
  \bibfield  {author} {\bibinfo {author} {\bibfnamefont {M.~G.}\ \bibnamefont
  {Allingham}}\ and\ \bibinfo {author} {\bibfnamefont {A.}~\bibnamefont
  {Sandmo}},\ }\bibfield  {title} {\bibinfo {title} {{Income tax evasion: a
  theoretical analysis}},\ }\href
  {https://doi.org/10.1016/0047-2727(72)90010-2} {\bibfield  {journal}
  {\bibinfo  {journal} {J. Public Econ.}\ }\textbf {\bibinfo {volume} {1}},\
  \bibinfo {pages} {323} (\bibinfo {year} {1972})}\BibitemShut {NoStop}%
\bibitem [{\citenamefont {Yitzhaki}(1974)}]{Yitzhaki1974}%
  \BibitemOpen
  \bibfield  {author} {\bibinfo {author} {\bibfnamefont {S.}~\bibnamefont
  {Yitzhaki}},\ }\bibfield  {title} {\bibinfo {title} {{Income tax evasion: A
  theoretical analysis}},\ }\href
  {https://doi.org/10.1016/0047-2727(74)90037-1} {\bibfield  {journal}
  {\bibinfo  {journal} {J. Public Econ.}\ }\textbf {\bibinfo {volume} {3}},\
  \bibinfo {pages} {201} (\bibinfo {year} {1974})}\BibitemShut {NoStop}%
\bibitem [{\citenamefont {Bosco}\ and\ \citenamefont
  {Mittone}(1997)}]{Bosco1997}%
  \BibitemOpen
  \bibfield  {author} {\bibinfo {author} {\bibfnamefont {L.}~\bibnamefont
  {Bosco}}\ and\ \bibinfo {author} {\bibfnamefont {L.}~\bibnamefont
  {Mittone}},\ }\bibfield  {title} {\bibinfo {title} {{Tax Evasion and Moral
  Constraints: some Experimental Evidence}},\ }\href
  {https://doi.org/10.1111/1467-6435.00018} {\bibfield  {journal} {\bibinfo
  {journal} {Kyklos}\ }\textbf {\bibinfo {volume} {50}},\ \bibinfo {pages}
  {297} (\bibinfo {year} {1997})}\BibitemShut {NoStop}%
\bibitem [{\citenamefont {Mittone}(2006)}]{Mittone2006}%
  \BibitemOpen
  \bibfield  {author} {\bibinfo {author} {\bibfnamefont {L.}~\bibnamefont
  {Mittone}},\ }\bibfield  {title} {\bibinfo {title} {{Dynamic behaviour in tax
  evasion: An experimental approach}},\ }\href
  {https://doi.org/10.1016/j.socec.2005.11.065} {\bibfield  {journal} {\bibinfo
   {journal} {J. Socio. Econ.}\ }\textbf {\bibinfo {volume} {35}},\ \bibinfo
  {pages} {813} (\bibinfo {year} {2006})}\BibitemShut {NoStop}%
\bibitem [{\citenamefont {Hokamp}\ and\ \citenamefont
  {Pickhardt}(2010)}]{Hokamp2010a}%
  \BibitemOpen
  \bibfield  {author} {\bibinfo {author} {\bibfnamefont {S.}~\bibnamefont
  {Hokamp}}\ and\ \bibinfo {author} {\bibfnamefont {M.}~\bibnamefont
  {Pickhardt}},\ }\bibfield  {title} {\bibinfo {title} {{Income Tax Evasion in
  a Society of Heterogeneous Agents – Evidence from an Agent-based Model}},\
  }\href {https://doi.org/10.1080/10168737.2010.525994} {\bibfield  {journal}
  {\bibinfo  {journal} {Int. Econ. J.}\ }\textbf {\bibinfo {volume} {24}},\
  \bibinfo {pages} {541} (\bibinfo {year} {2010})}\BibitemShut {NoStop}%
\bibitem [{\citenamefont {Garrido}\ and\ \citenamefont
  {Mittone}(2013)}]{Garrido2013}%
  \BibitemOpen
  \bibfield  {author} {\bibinfo {author} {\bibfnamefont {N.}~\bibnamefont
  {Garrido}}\ and\ \bibinfo {author} {\bibfnamefont {L.}~\bibnamefont
  {Mittone}},\ }\bibfield  {title} {\bibinfo {title} {{An agent based model for
  studying optimal tax collection policy using experimental data: The cases of
  Chile and Italy}},\ }\href {https://doi.org/10.1016/j.socec.2012.11.002}
  {\bibfield  {journal} {\bibinfo  {journal} {J. Socio. Econ.}\ }\textbf
  {\bibinfo {volume} {42}},\ \bibinfo {pages} {24} (\bibinfo {year}
  {2013})}\BibitemShut {NoStop}%
\bibitem [{\citenamefont {Hokamp}\ and\ \citenamefont
  {Seibold}(2014)}]{Hokamp2014}%
  \BibitemOpen
  \bibfield  {author} {\bibinfo {author} {\bibfnamefont {S.}~\bibnamefont
  {Hokamp}}\ and\ \bibinfo {author} {\bibfnamefont {G.}~\bibnamefont
  {Seibold}},\ }\bibfield  {title} {\bibinfo {title} {{How Much Rationality
  Tolerates the Shadow Economy? – An Agent-Based Econophysics Approach}},\
  }in\ \href {https://doi.org/10.1007/978-3-642-39829-2_11} {\emph {\bibinfo
  {booktitle} {Adv. Intell. Syst. Comput.}}}\ (\bibinfo {year} {2014})\ pp.\
  \bibinfo {pages} {119--128}\BibitemShut {NoStop}%
\bibitem [{\citenamefont {Koehler}\ \emph {et~al.}(2018)\citenamefont
  {Koehler}, \citenamefont {Michel}, \citenamefont {Slater}, \citenamefont
  {Harvey}, \citenamefont {Andrei},\ and\ \citenamefont {Comer}}]{Koehler2018}%
  \BibitemOpen
  \bibfield  {author} {\bibinfo {author} {\bibfnamefont {M.}~\bibnamefont
  {Koehler}}, \bibinfo {author} {\bibfnamefont {S.}~\bibnamefont {Michel}},
  \bibinfo {author} {\bibfnamefont {D.}~\bibnamefont {Slater}}, \bibinfo
  {author} {\bibfnamefont {C.}~\bibnamefont {Harvey}}, \bibinfo {author}
  {\bibfnamefont {A.}~\bibnamefont {Andrei}},\ and\ \bibinfo {author}
  {\bibfnamefont {K.}~\bibnamefont {Comer}},\ }\bibfield  {title} {\bibinfo
  {title} {{Investigating the Effects of Network Structures in Massive
  Agent-Based Models of Tax Evasion}},\ }in\ \href
  {https://doi.org/10.1002/9781119155713.ch8} {\emph {\bibinfo {booktitle}
  {Agent-based Model. Tax Evas.}}}\ (\bibinfo  {publisher} {John Wiley {\&}
  Sons, Ltd},\ \bibinfo {address} {Chichester, UK},\ \bibinfo {year} {2018})\
  pp.\ \bibinfo {pages} {225--253}\BibitemShut {NoStop}%
\bibitem [{\citenamefont {{Garcia Alvarado}}(2019)}]{GarciaAlvarado2019}%
  \BibitemOpen
  \bibfield  {author} {\bibinfo {author} {\bibfnamefont {F.}~\bibnamefont
  {{Garcia Alvarado}}},\ }\bibfield  {title} {\bibinfo {title} {{Network
  Effects in an Agent-Based Model of Tax Evasion with Social Influence}},\ }in\
  \href {https://doi.org/10.1007/978-3-030-24209-1_7} {\emph {\bibinfo
  {booktitle} {Lect. Notes Comput. Sci.}}}\ (\bibinfo {year} {2019})\ pp.\
  \bibinfo {pages} {78--89}\BibitemShut {NoStop}%
\bibitem [{\citenamefont {Zaklan}\ \emph {et~al.}(2008)\citenamefont {Zaklan},
  \citenamefont {Lima},\ and\ \citenamefont {Westerhoff}}]{Zaklan2008}%
  \BibitemOpen
  \bibfield  {author} {\bibinfo {author} {\bibfnamefont {G.}~\bibnamefont
  {Zaklan}}, \bibinfo {author} {\bibfnamefont {F.}~\bibnamefont {Lima}},\ and\
  \bibinfo {author} {\bibfnamefont {F.}~\bibnamefont {Westerhoff}},\ }\bibfield
   {title} {\bibinfo {title} {{Controlling tax evasion fluctuations}},\ }\href
  {https://doi.org/10.1016/j.physa.2008.06.036} {\bibfield  {journal} {\bibinfo
   {journal} {Phys. A Stat. Mech. its Appl.}\ }\textbf {\bibinfo {volume}
  {387}},\ \bibinfo {pages} {5857} (\bibinfo {year} {2008})}\BibitemShut
  {NoStop}%
\bibitem [{\citenamefont {LIMA}\ and\ \citenamefont {ZAKLAN}(2008)}]{Lima2008}%
  \BibitemOpen
  \bibfield  {author} {\bibinfo {author} {\bibfnamefont {F.~W.~S.}\
  \bibnamefont {LIMA}}\ and\ \bibinfo {author} {\bibfnamefont {G.}~\bibnamefont
  {ZAKLAN}},\ }\bibfield  {title} {\bibinfo {title} {{A MULTI-AGENT-BASED
  APPROACH TO TAX MORALE}},\ }\href {https://doi.org/10.1142/S0129183108013357}
  {\bibfield  {journal} {\bibinfo  {journal} {Int. J. Mod. Phys. C}\ }\textbf
  {\bibinfo {volume} {19}},\ \bibinfo {pages} {1797} (\bibinfo {year}
  {2008})},\ \Eprint {https://arxiv.org/abs/0806.0344} {arXiv:0806.0344}
  \BibitemShut {NoStop}%
\bibitem [{\citenamefont {Zaklan}\ \emph {et~al.}(2009)\citenamefont {Zaklan},
  \citenamefont {Westerhoff},\ and\ \citenamefont {Stauffer}}]{Zaklan2009}%
  \BibitemOpen
  \bibfield  {author} {\bibinfo {author} {\bibfnamefont {G.}~\bibnamefont
  {Zaklan}}, \bibinfo {author} {\bibfnamefont {F.}~\bibnamefont {Westerhoff}},\
  and\ \bibinfo {author} {\bibfnamefont {D.}~\bibnamefont {Stauffer}},\
  }\bibfield  {title} {\bibinfo {title} {{Analysing tax evasion dynamics via
  the Ising model}},\ }\href {https://doi.org/10.1007/s11403-008-0043-5}
  {\bibfield  {journal} {\bibinfo  {journal} {J. Econ. Interact. Coord.}\
  }\textbf {\bibinfo {volume} {4}},\ \bibinfo {pages} {1} (\bibinfo {year}
  {2009})},\ \Eprint {https://arxiv.org/abs/0801.2980} {arXiv:0801.2980}
  \BibitemShut {NoStop}%
\bibitem [{\citenamefont {Ising}(1925)}]{Ising1925}%
  \BibitemOpen
  \bibfield  {author} {\bibinfo {author} {\bibfnamefont {E.}~\bibnamefont
  {Ising}},\ }\bibfield  {title} {\bibinfo {title} {{Beitrag zur Theorie des
  Ferromagnetismus}},\ }\href {https://doi.org/10.1007/BF02980577} {\bibfield
  {journal} {\bibinfo  {journal} {Zeitschrift f{\"{u}}r Phys.}\ }\textbf
  {\bibinfo {volume} {31}},\ \bibinfo {pages} {253} (\bibinfo {year}
  {1925})}\BibitemShut {NoStop}%
\bibitem [{\citenamefont {Lima}(2012)}]{Lima2012a}%
  \BibitemOpen
  \bibfield  {author} {\bibinfo {author} {\bibfnamefont {F.~W.~S.}\
  \bibnamefont {Lima}},\ }\bibfield  {title} {\bibinfo {title} {{Tax Evasion
  and Nonequilibrium Model on Apollonian Networks}},\ }\href
  {https://doi.org/10.1142/S0129183112500799} {\bibfield  {journal} {\bibinfo
  {journal} {Int. J. Mod. Phys. C}\ }\textbf {\bibinfo {volume} {23}},\
  \bibinfo {pages} {1250079} (\bibinfo {year} {2012})}\BibitemShut {NoStop}%
\bibitem [{\citenamefont {LIMA}(2012)}]{Lima2012}%
  \BibitemOpen
  \bibfield  {author} {\bibinfo {author} {\bibfnamefont {F.~W.~S.}\
  \bibnamefont {LIMA}},\ }\bibfield  {title} {\bibinfo {title} {{Tax Evasion
  Dynamics and Zaklan Models on Opinion-Dependent Network}},\ }\href
  {https://doi.org/10.1142/S0129183112500477} {\bibfield  {journal} {\bibinfo
  {journal} {Int. J. Mod. Phys. C}\ }\textbf {\bibinfo {volume} {23}},\
  \bibinfo {pages} {1250047} (\bibinfo {year} {2012})},\ \Eprint
  {https://arxiv.org/abs/1204.0386} {arXiv:1204.0386} \BibitemShut {NoStop}%
\bibitem [{\citenamefont {Lima}(2015{\natexlab{a}})}]{Lima2015}%
  \BibitemOpen
  \bibfield  {author} {\bibinfo {author} {\bibfnamefont {F.~W.~S.}\
  \bibnamefont {Lima}},\ }\bibfield  {title} {\bibinfo {title} {{Tax Evasion
  Dynamics via Non-Equilibrium Model on Complex Networks}},\ }\href
  {https://doi.org/10.4236/tel.2015.56089} {\bibfield  {journal} {\bibinfo
  {journal} {Theor. Econ. Lett.}\ }\textbf {\bibinfo {volume} {05}},\ \bibinfo
  {pages} {775} (\bibinfo {year} {2015}{\natexlab{a}})}\BibitemShut {NoStop}%
\bibitem [{\citenamefont {Lima}(2015{\natexlab{b}})}]{Lima2015a}%
  \BibitemOpen
  \bibfield  {author} {\bibinfo {author} {\bibfnamefont {F.~W.~S.}\
  \bibnamefont {Lima}},\ }\bibfield  {title} {\bibinfo {title} {{Tax evasion
  dynamics and nonequilibrium Zaklan model with heterogeneous agents on square
  lattice}},\ }\href {https://doi.org/10.1142/S0129183115500357} {\bibfield
  {journal} {\bibinfo  {journal} {Int. J. Mod. Phys. C}\ }\textbf {\bibinfo
  {volume} {26}},\ \bibinfo {pages} {1550035} (\bibinfo {year}
  {2015}{\natexlab{b}})}\BibitemShut {NoStop}%
\bibitem [{\citenamefont {Seibold}\ and\ \citenamefont
  {Pickhardt}(2013)}]{Seibold2013b}%
  \BibitemOpen
  \bibfield  {author} {\bibinfo {author} {\bibfnamefont {G.}~\bibnamefont
  {Seibold}}\ and\ \bibinfo {author} {\bibfnamefont {M.}~\bibnamefont
  {Pickhardt}},\ }\bibfield  {title} {\bibinfo {title} {{Lapse of time effects
  on tax evasion in an agent-based econophysics model}},\ }\href
  {https://doi.org/10.1016/j.physa.2013.01.016} {\bibfield  {journal} {\bibinfo
   {journal} {Phys. A Stat. Mech. its Appl.}\ }\textbf {\bibinfo {volume}
  {392}},\ \bibinfo {pages} {2079} (\bibinfo {year} {2013})}\BibitemShut
  {NoStop}%
\bibitem [{\citenamefont {Pickhardt}\ and\ \citenamefont
  {Seibold}(2014)}]{Pickhardt2014a}%
  \BibitemOpen
  \bibfield  {author} {\bibinfo {author} {\bibfnamefont {M.}~\bibnamefont
  {Pickhardt}}\ and\ \bibinfo {author} {\bibfnamefont {G.}~\bibnamefont
  {Seibold}},\ }\bibfield  {title} {\bibinfo {title} {{Income tax evasion
  dynamics: Evidence from an agent-based econophysics model}},\ }\href
  {https://doi.org/10.1016/j.joep.2013.01.011} {\bibfield  {journal} {\bibinfo
  {journal} {J. Econ. Psychol.}\ }\textbf {\bibinfo {volume} {40}},\ \bibinfo
  {pages} {147} (\bibinfo {year} {2014})},\ \Eprint
  {https://arxiv.org/abs/1112.0233} {arXiv:1112.0233} \BibitemShut {NoStop}%
\bibitem [{\citenamefont {Bazart}\ \emph {et~al.}(2016)\citenamefont {Bazart},
  \citenamefont {Bonein}, \citenamefont {Hokamp},\ and\ \citenamefont
  {Seibold}}]{Bazart2016}%
  \BibitemOpen
  \bibfield  {author} {\bibinfo {author} {\bibfnamefont {C.}~\bibnamefont
  {Bazart}}, \bibinfo {author} {\bibfnamefont {A.}~\bibnamefont {Bonein}},
  \bibinfo {author} {\bibfnamefont {S.}~\bibnamefont {Hokamp}},\ and\ \bibinfo
  {author} {\bibfnamefont {G.}~\bibnamefont {Seibold}},\ }\bibfield  {title}
  {\bibinfo {title} {{Bahavioural Economics and Tax Evasion: Calibrating an
  Agent-based Econophysics Model with Experimental Tax Compliance Data}},\
  }\href@noop {} {\bibfield  {journal} {\bibinfo  {journal} {J. Tax Adm.}\ }
  (\bibinfo {year} {2016})}\BibitemShut {NoStop}%
\bibitem [{\citenamefont {Berger}\ \emph {et~al.}(2020)\citenamefont {Berger},
  \citenamefont {Hokamp},\ and\ \citenamefont {Seibold}}]{Berger2020}%
  \BibitemOpen
  \bibfield  {author} {\bibinfo {author} {\bibfnamefont {W.}~\bibnamefont
  {Berger}}, \bibinfo {author} {\bibfnamefont {S.}~\bibnamefont {Hokamp}},\
  and\ \bibinfo {author} {\bibfnamefont {G.}~\bibnamefont {Seibold}},\
  }\bibfield  {title} {\bibinfo {title} {{Dynamic behavioural changes in an
  agent-based econophysics tax compliance model: bomb crater versus target
  effects and efficient audit strategies}},\ }\bibfield  {journal} {\bibinfo
  {journal} {J. Public Financ. Public Choice}\ }\href
  {https://doi.org/10.1332/251569120X15840237292628}
  {10.1332/251569120X15840237292628} (\bibinfo {year} {2020})\BibitemShut
  {NoStop}%
\bibitem [{\citenamefont {Jimenez}\ and\ \citenamefont
  {Jacinto}(2010)}]{Jimenez2010}%
  \BibitemOpen
  \bibfield  {author} {\bibinfo {author} {\bibfnamefont {O.~P.}\ \bibnamefont
  {Jimenez}}\ and\ \bibinfo {author} {\bibfnamefont {R.~P.}\ \bibnamefont
  {Jacinto}},\ }\bibfield  {title} {\bibinfo {title} {Evasión de impuestos
  nacionales en colombia: Años 2001 - 2009},\ }\href@noop {} {\bibfield
  {journal} {\bibinfo  {journal} {Revista Facultad de Ciencias Económicas,
  Universidad Nueva Granada.}\ }\textbf {\bibinfo {volume} {2}},\ \bibinfo
  {pages} {177 } (\bibinfo {year} {2010})}\BibitemShut {NoStop}%
\bibitem [{\citenamefont {Rodriguez-Cuervo}(2018)}]{Cuervo2018}%
  \BibitemOpen
  \bibfield  {author} {\bibinfo {author} {\bibfnamefont {J.}~\bibnamefont
  {Rodriguez-Cuervo}},\ }\bibfield  {title} {\bibinfo {title} {Análisis y
  estimación de la evasión y elusión de impuestos en colombia durante el
  periodo 1997 -2017, e identificación de los principales cambios tributarios
  generados para combatirlos},\ }\href
  {https://repositorio.unal.edu.co/handle/unal/68799?show=full} {\bibfield
  {journal} {\bibinfo  {journal} {Trabajo de Grado, Maestría en
  Administración de Empresas, Universidad Nacional de Colombia.}\ } (\bibinfo
  {year} {2018})}\BibitemShut {NoStop}%
\bibitem [{\citenamefont {van Rossum}\ and\ \citenamefont
  {Drake}(2009)}]{VanRossum2009}%
  \BibitemOpen
  \bibfield  {author} {\bibinfo {author} {\bibfnamefont {G.}~\bibnamefont {van
  Rossum}}\ and\ \bibinfo {author} {\bibfnamefont {F.~L.}\ \bibnamefont
  {Drake}},\ }\href@noop {} {\emph {\bibinfo {title} {Scotts Valley, CA}}}\
  (\bibinfo {year} {2009})\BibitemShut {NoStop}%
\bibitem [{\citenamefont {Krauth}(2006)}]{Krauth2006}%
  \BibitemOpen
  \bibfield  {author} {\bibinfo {author} {\bibfnamefont {W.}~\bibnamefont
  {Krauth}},\ }\href@noop {} {\emph {\bibinfo {title} {Statistical Mechanics -
  Algor. and Comput.}}}\ (\bibinfo  {publisher} {Oxford University Press},\
  \bibinfo {year} {2006})\BibitemShut {NoStop}%
\bibitem [{\citenamefont {J{\c{e}}drzejewski}\ \emph
  {et~al.}(2017)\citenamefont {J{\c{e}}drzejewski}, \citenamefont {Chmiel},\
  and\ \citenamefont {Sznajd-Weron}}]{Jedrzejewski2017}%
  \BibitemOpen
  \bibfield  {author} {\bibinfo {author} {\bibfnamefont {A.}~\bibnamefont
  {J{\c{e}}drzejewski}}, \bibinfo {author} {\bibfnamefont {A.}~\bibnamefont
  {Chmiel}},\ and\ \bibinfo {author} {\bibfnamefont {K.}~\bibnamefont
  {Sznajd-Weron}},\ }\bibfield  {title} {\bibinfo {title} {{Kinetic Ising
  models with various single-spin-flip dynamics on quenched and annealed random
  regular graphs}},\ }\href {https://doi.org/10.1103/PhysRevE.96.012132}
  {\bibfield  {journal} {\bibinfo  {journal} {Phys. Rev. E}\ }\textbf {\bibinfo
  {volume} {96}},\ \bibinfo {pages} {012132} (\bibinfo {year} {2017})},\
  \Eprint {https://arxiv.org/abs/1703.03602} {arXiv:1703.03602} \BibitemShut
  {NoStop}%
\bibitem [{\citenamefont {Crokidakis}(2014)}]{Crokidakis2014a}%
  \BibitemOpen
  \bibfield  {author} {\bibinfo {author} {\bibfnamefont {N.}~\bibnamefont
  {Crokidakis}},\ }\bibfield  {title} {\bibinfo {title} {{A three-state kinetic
  agent-based model to analyze tax evasion dynamics}},\ }\href
  {https://doi.org/10.1016/j.physa.2014.07.056} {\bibfield  {journal} {\bibinfo
   {journal} {Phys. A Stat. Mech. its Appl.}\ }\textbf {\bibinfo {volume}
  {414}},\ \bibinfo {pages} {321} (\bibinfo {year} {2014})},\ \Eprint
  {https://arxiv.org/abs/1407.5220} {arXiv:1407.5220} \BibitemShut {NoStop}%
\end{thebibliography}%

\end{document}